\documentclass{aa}

\usepackage{graphicx}
\usepackage{natbib}
\usepackage{booktabs}
\bibpunct{(}{)}{;}{a}{}{,}
\usepackage{txfonts}
\usepackage[colorlinks=true,citecolor=blue,linkcolor=blue]{hyperref}
\usepackage{multicol}
\usepackage{subcaption}
\newcommand{\rt}[1]{{#1}}

\newcommand{\microns}{\mu \mathrm{m}}
\newcommand{\amin}{a_{\mathrm{min}}}
\newcommand{\amax}{a_{\mathrm{max}}}
\newcommand{\Rin}{R_{\mathrm{in}}}
\newcommand{\Rout}{R_{\mathrm{out}}}
\newcommand{\rin}{r_{\mathrm{in}}}
\newcommand{\rout}{r_{\mathrm{out}}}
\newcommand{\mcmax}{\texttt{MCMax3D}}
\newcommand{\Rtap}{R_{\mathrm{tap}}}
\newcommand{\Mdust}{M_{\mathrm{dust}}}
\newcommand{\dd}{\mathrm{d}}
\newcommand{\Qphi}{Q_{\phi}}
\newcommand{\Uphi}{U_{\phi}}
\newcommand{\Mjup}{M_{\mathrm{Jup}}}
\newcommand{\Rjup}{R_{\mathrm{Jup}}}
\newcommand{\Mearth}{M_{\oplus}}
\newcommand{\Lsun}{L_{\odot}}
\newcommand{\Halpha}{\mathrm{H}\alpha}

\newcommand{\Rhill}{r_{\mathrm{Hill}}}
\newcommand{\kabs}{\kappa_{\nu}^{\mathrm{abs}}}
\newcommand{\kb}{k_{\mathrm{B}}}
\newcommand{\Tave}{\langle T \rangle}

\newcommand{\amean}{\langle a^3 \rangle^{1/3}}

\begin{document}

   \title{Self-consistent modelling of the dust component in protoplanetary and circumplanetary disks: The case of PDS 70}
   %\subtitle{I. Overviewing the $\kappa$-mechanism}

   \author{B. Portilla-Revelo \inst{1},
          I. Kamp \inst{1}, Ch. Rab \inst{1} \fnmsep \inst{6} \fnmsep \inst{7}, E. F. van Dishoeck \inst{5} \fnmsep \inst{6}, M. Keppler \inst{4}, M. Min \inst{2} \fnmsep \inst{3} and
          G.A. Muro-Arena \inst{2}% \thanks{Just to show the usage
          %of the elements in the author field}
          }
    \authorrunning{B. Portilla Revelo et al.}
    \titlerunning{Self-consistent modelling of the dust component in protoplanetary and circumplanetary disks}

   \institute{Kapteyn Astronomical Institute, University of Groningen,
              PO Box 800, 9700 AV Groningen, The Netherlands \\
              \email{bportilla@astro.rug.nl}
        \and
            Anton Pannekoek Institute for Astronomy, University of Amsterdam, Science Park 904, 1098 XH Amsterdam, The Netherlands
        \and
            SRON Netherlands Institute for Space Research, Sorbonnelaan 2, 3584 CA Utrecht, The Netherlands
        \and
            Max Planck Institute for Astronomy, Königstuhl 17, 69117 Heidelberg, Germany
        \and
            Leiden Observatory, Leiden University, PO Box 9513, 2300 RA Leiden, The Netherlands
        \and
            Max-Planck-Institut für Extraterrestrische Physik, Giessenbachstrasse1, 85748 Garching, Germany
        \and
            University Observatory, Faculty of Physics, Ludwig-Maximilians-Universität München, Scheinerstr. 1, 81679 Munich, Germany
             }

   \date{Accepted November 9, 2021}
   %\date{}

% \abstract{}{}{}{}{}
% 5 {} token are mandatory


  \abstract
   {Direct observations of young stellar objects are important to test established theories of planet formation. PDS 70 is one of the few cases where robust evidence favours the presence of two planetary mass companions inside the gap of the transition disk. Those planets are believed to be going through the last stages of accretion from the protoplanetary disk, a process likely mediated by a circumplanetary disk (CPD).}
   {We aim to develop a three-dimensional radiative transfer model for the dust component of the PDS 70 system which reproduces the system's global features observed at two different wavelengths: $855\, \microns$ with the Atacama large millimeter/submillimeter array (ALMA) and $1.25\, \microns$ with the Spectro-polarimetric high-contrast exoplanet research instrument at the Very large telescope (VLT/SPHERE). We use this model to \rt{investigate} the physical properties of the planetary companion PDS 70 c and its potential circumplanetary disk.}
   {We base our modelling process on published ALMA and VLT/SPHERE observations of the dust continuum emission at $855 \, \microns$ and $1.25 \, \microns$, respectively. We selected initial values for the physical properties of the planet and CPD through appropriate assumptions about the nature and evolutionary stage of the object. We modified the properties of the protoplanetary disk iteratively until the predictions retrieved from the model were consistent with both data sets. Simulations were carried out with the three-dimensional radiative transfer code \mcmax.}
   {We provide a model that jointly explains the global features of the PDS 70 system seen in sub-millimetre and polarised-scattered light. Our model suggests that spatial segregation of dust grains is present in the protoplanetary disk. The sub-millimetre modelling of the PDS 70 c source favours the presence of \rt{an optically thick} CPD and \rt{places an upper limit on its dust mass of $0.7\, \Mearth$}. Furthermore, analysis of the thermal structure of the CPD demonstrates that the planet luminosity is the dominant heating mechanism of dust grains inside 0.6 au from the planet, while heating by stellar photons dominates at larger planetocentric distances.}
   {A CPD surrounding the planetary-mass companion PDS 70 c is a plausible scenario to explain the substructure observed with ALMA. The heating feedback from the protoplanetary disk has an non-negligible effect on the equilibrium temperature of dust grains in the outskirts of the CPD. The connection between the CPD properties and the planet mass still depends on a series of key assumptions. Further observations with high spatial and spectral resolution also for the gas component of the CPD are required to break the degeneracy between the properties of the planet and the disk.}

   \keywords{Protoplanetary disks ---
   Planets and satellites: formation ---
   Stars: individual: PDS 70 ---
   Submillimeter: planetary systems ---
   Infrared: planetary systems ---
   Methods: numerical
   }

   \maketitle
%
%-------------------------------------------------------------------

\section{Introduction}
    The late stages in the formation of gas giant planets are characterised by the presence of a disk made of dust and gas which surrounds and feeds the forming planet. Such a structure receives the name of circumplanetary disk (CPD) \citep{Lubow1999,Kley1999}, and it is where the formation of large and regular moons is thought to take place \citep{Canup2002}. Hydrodynamical simulations have shown that the emergence of these structures is common both in the core accretion and in the gravitational instability models of planetary formation \citep{Szulagyi2017}. Furthermore, the architectures of large moon systems, such as those observed around Jupiter and Saturn, are consistent with the constraints imposed by satellite formation via CPDs \citep{Sasaki2010}. However, direct and conclusive observational evidence of CPDs remains elusive.
    The sub-millimetre emission from the surroundings of planet-forming regions is an important window to scrutinise the complexity of planetary formation. Dust particles in the CPD are heated by the visible and near-infrared (NIR) radiation received from the star and the planet. This radiation is then re-emitted at longer wavelengths, making these disks shine in the far-infrared and sub-millimetre regimes \citep{Draine2010,Armitage2010}.

    The first report of a potential CPD candidate was done by \cite{Welch2004}. Observations made at $1.4$ mm of the T-Tauri star HL Tau showed a localised enhancement of the surface density at a distance of $70$ AU from the central star. The feature was interpreted as dust emission associated with a protoplanet candidate, but independent measurements by \cite{Carrasco2009} at $\lambda=7 \, \mathrm{mm}$ showed that no emission above $3\sigma$ is coming from the location of the potential candidate. They suggested that the signal was probably due to free-free rather than thermal emission making the real nature of the overdensity arguable.

    \cite{Kraus2012} studied the LkCa 15 system known for hosting a transition disk with a gap of width $\sim 55$ au. They reported the direct imaging of a potential companion candidate close to the centre of the gap which shows an extended structure when observed in $L'$ and $K'$ bands, which is not consistent with the expected footprint of an unresolved point source, such as a planet. The observed structure was interpreted as a forming gas giant planet surrounded by circumplanetary material. Very large array (VLA) observations of the 7 mm continuum emission from the same system were carried out by \cite{Isella2014}, but those did not detect any emission coming from the position of the potential companion. Similarly, more recent  observations carried out with the Atacama large millimeter/submillimeter array (ALMA) could not detect with enough statistical significance any localised sub-millimetre signal within the cavity of LkCa 15  \citep{Facchini2020}.

    These non-confirmed CPD candidates around HL Tau and LkCa 15 illustrate how challenging it is to achieve a non-ambiguous detection of CPDs embedded in dust-depleted cavities of protoplanetary disks. Modelling and high resolution observations at multiple wavelengths are required to understand the environment around forming giant planets.

    An interesting candidate for the detection of CPDs is PDS 70 and its protoplanetary disk. PDS 70 is a $0.76 \, M_{\odot}$ T-Tauri star \citep{Muller2018} located at a $113.5$ pc distance in the Centaurus constellation \citep{Prusti2016,Brown2018}. The first observation made in scattered light in the $K_s$-band was able to mask the stellar brightness, keeping an angular resolution high enough to provide evidence for the presence of a protoplanetary disk around the star \citep{Riaud2006}. Observations in bands $H$ and $L'$ showed that this disk has a cavity with a radial extension of $\sim 70$ au \citep{Hashimoto2012}. Based on the observational data, and after a process of radiative transfer modelling, the authors proposed that the cavity could be the result of the interaction between the disk and a number of companions of a sub-stellar or planetary nature. Their results also led them to propose a first estimate for the upper mass limit of the companion bodies which was about $50 \, \Mjup$. This was the first time the existence of at least one forming giant planet around PDS 70 was hypothesised.

    \cite{Keppler2018} reported the discovery of a planetary mass companion inside the cavity at a  distance of $\sim 22$ au based on near-infrared observations. In addition, this work was the first to report polarised light coming from an inner disk. \cite{Haffert2019} provided an independent confirmation of PDS 70 b via $\Halpha$ observations and also announced the presence of an additional $\Halpha$ source inside the cavity, labelled as PDS 70 c. The latter observation was confirmed at longer wavelengths by \cite{Isella2019} who claimed that the observed sub-millimetre flux could be explained by the presence of circumplanetary material around the forming planet PDS 70 c.

    The signal coming from the PDS 70 c location is spatially unresolved and its closeness to the outer protoplanetary dust ring \citep{Keppler2019,Mesa2019} makes it difficult to isolate the observation of the planet-CPD system from the protoplanetary disk. The ability to separate the individual contributions from the planet, CPD, and protoplanetary disk to the total sub-millimetre flux is essential. Indeed, it should be noted that the CPD brightness can be greater than the planet's, at least by one order of magnitude \citep{Szulagyi2019}. Therefore, using the combined luminosity to estimate, for example, the planetary mass can lead to an overestimate of the real value by a similar extent. \rt{The recent high resolution continuum observations of PDS 70 reported in \cite{Benisty2021} allow one to separate the CPD emission from the outer disk and placed a robust constraint on the CPD size.}

    To conclude, although there is robust evidence to support the presence of two gap-carving planets around PDS 70, efforts are still required to fully characterise the architecture of the system. Recent works have put important constraints on both the orbital and bulk properties of the two planets, especially for PDS 70 b which is located well inside the gap (e.g. \citealt{Muller2018,Wang2020,Wang2021}). For PDS 70 c, however, the determination of those properties is more challenging and further analysis of the circumplanetary material and its impact on the observations is required. This work aims to shed light on these questions.

    \rt{This paper is organised as follows. In Sect. \ref{sec:rt_model} we present a two-dimensional radiative transfer model to explain self-consistently the sub-millimetre and NIR continuum emission from the PDS 70 disk detected at $855 \, \microns$ and at $1.25 \, \microns$, respectively. This reference model does not include any planetary mass companions, but paves the way to evaluate the effect that an additional source mimicking PDS 70 c has on the observations. In Sect. \ref{sec:including_a_planet} we extend the model to a three-dimensional case by including a planetary companion at the predicted location of PDS 70 c and we show the effects this additional source would have in the synthesised ALMA image when the planet is simulated with and without a CPD around it. In Sect \ref{sec:discusion} we derive constraints for the amount of dust in the CPD and discuss our results in light of the work of \cite{Benisty2021}. We devote special attention to the thermal structure of the CPD to constrain the relative influence of the star and the planet in the heating of the dust grains. We summarise our findings in Sect. \ref{sec:conclusions}.}

\section {Multiwavelength radiative transfer modelling}
\label{sec:rt_model}

In this section we describe the process followed to find a suitable model to reproduce three observables of the PDS 70 system: the ALMA image observed at $855 \, \microns$, the VLT/SPHERE image at $1.25 \, \microns$ in polarised and scattered light and the photometry of the source. To our knowledge, this is the first attempt to develop a self-consistent radiative transfer model for these particular data sets. For a description of the observation and data reduction process, we refer the reader to the works of \cite{Isella2019}, \cite{Keppler2018} and references therein.

Our strategy consists in first modelling the ALMA data. This allows us to find the modulated profile for the surface density of a homogeneous distribution of dust particles. This profile is then used as an input parameter to model the $\Qphi$ component of the polarised-scattered emission. We show that this step requires to abandon the idea of a homogeneous distribution of dust particles and change it for a radial segregation of particle sizes throughout the disk that also shows a differential stratification in the vertical direction. A final tuning of some geometrical parameters is also required in order to fit the modelled spectral energy distribution with the observed photometry. We used the three-dimensional version of the continuum radiative transfer code \texttt{MCMax} \citep{Min2009}.

\begin{figure}[h!]
\centering
\includegraphics[width=1.0\linewidth]{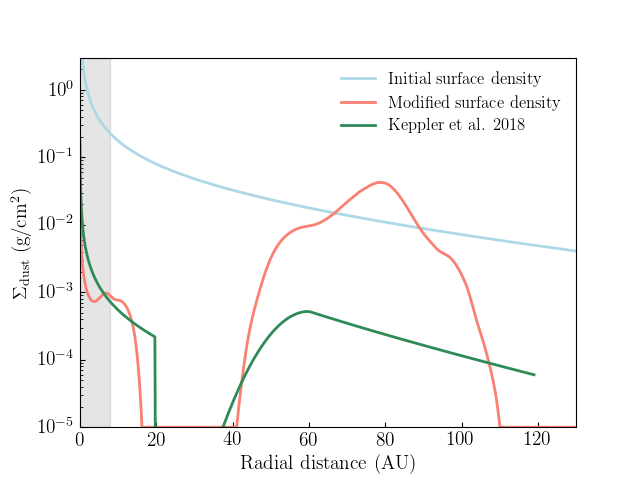}
\caption{\small Surface density profiles before and after the iterative procedure described in Sect.~\ref{sect:alma_modeling}. \rt{The modified profile provides the best fit to the ALMA data}. The grey area depicts the spatially unresolved inner disk due to the ALMA beam size.}
\label{fig:density_profiles}
\end{figure}

\subsection{Modelling the dust continuum emission}
\label{sect:alma_modeling}
Thermal emission is dominated by the large dust particles which tend to settle down into the midplane of the disk. The radial surface density profile is a good proxy to the thermal dust emission as it traces how the particles are distributed along the disk. We started our modelling with a simple prescription for the surface density profile which is assumed axially symmetric:

\begin{equation}
\label{eq:surface_density}
\Sigma_{\mathrm{dust}}(r)=\Sigma_0 \bigg(\frac{\Rtap}{r}\bigg)^\epsilon\exp\Bigg[-\bigg(\frac{r}{\Rtap}\bigg)^{2-\gamma}\Bigg],
\end{equation}

\noindent where $\Rtap$ is the tapering-off radius indicating the distance where the transition from a linear to an exponential decay of the surface density takes place. We use $\epsilon=\gamma=1$. The constant $\Sigma_0$ is determined by the total dust mass through the relation:

\begin{equation}
\label{eq:mdust}
\Mdust=\int\limits_{\Rin'}^{\Rout'} \int\limits_{0}^{2\pi} r \Sigma_{\mathrm{dust}}(r) \, \dd \phi \dd r,
\end{equation}

\noindent with $\Rin'$ and $\Rout'$ being the inner and outer limits of the protoplanetary disk, respectively. \rt{We adopt $\Rin'=0.04 \, \mathrm{au}$ following \cite{Keppler2018}. For the outer radius we assume $\Rout'=130 \, \mathrm{au}$ since the value of the azimuthally averaged flux at $855\, \microns$ oscillates around zero Jy for $r>120 \, \mathrm{au}$. A typical value for $\Rtap$ of $100 \, \mathrm{au}$ is taken from the T-Tauri reference model introduced by \cite{Woitke2016}.} The initial surface density profile is shown in Fig.~\ref{fig:density_profiles}. The disk has a flaring exponent $\beta$ and a scale height $H(r)$ with a characteristic value $H_0$ at a reference distance $r_0=\Rtap$:

\begin{equation}
\label{eq:hscale}
H(r)=H_0 \Bigg(\frac{r}{r_0}\Bigg)^{\beta}.
\end{equation}

We assume an initial distribution of dust particles with sizes between $0.05 \, \microns$ and $3$ mm with a size distribution $n(a)$ such that the number of particles with sizes between $a$ and $a+\dd a$ follows a power law:

\begin{equation}
\label{eq:size_distribution}
n(a) \dd a \propto a^{-p} \dd a,
\end{equation}

\noindent with $p$ being a positive exponent. The dust grains are a mixture of astronomical silicates and amorphous carbon with a porosity accounting for certain fraction of the total volume of the particle. This part of the modelling procedure used the DIANA standard dust opacities\footnote{\url{https://dianaproject.wp.st-andrews.ac.uk/data-results-downloads/fortran-package/}} \citep{Woitke2016}.

In \texttt{MCMax3D}, the treatment of dust settling follows the prescription given by \cite{Dubrulle1995}. The settling is parameterised through the turbulence parameter $\alpha$ \citep{Shakura1973,Min2012}; small values of $\alpha$ imply a high level of settling. For a fixed value of $\alpha$, the level of settling is higher for large and compact dust grains, although smaller particles can also follow the same trend but with a different efficiency. Therefore, the settling in our model is consistently implemented using a single parameter that affects both the small and the large grains. \rt{In section \ref{sec:vertical_segregation}, we discuss in more detail the parametrisation of the vertical dust distribution.}

\rt{For the vertical density structure, we use the following prescription:
\begin{equation}
\label{eq:gas_density}
\rho(r,z)=\frac{\Sigma(r)}{\sqrt{2\pi} \cdot H(r)}\exp{\bigg(-\frac{z^2}{2H(r)^2}\bigg)}\, ,
\end{equation}
where $\Sigma(r)=100 \times \Sigma_\mathrm{dust}(r)$ is the gas surface density and $H(r)$ is given by Eq. (\ref{eq:hscale})}.

We first ran a Monte Carlo simulation with the initial density profile shown in Fig.~\ref{fig:density_profiles}. The obtained disk structure was ray-traced at $\lambda=855 \, \microns$ and the resulting image was convolved with a Gaussian kernel that represents the ALMA beam of the observed data set. The beam size is $67\times 50$ mas ($\sim 8 \times 6$ au) and has a position angle of $61.5^\circ$ east of north. Then, we performed aperture photometry both in the observed and synthesised images to extract and compare the respective azimuthally averaged radial brightness profiles. In order to do this, we used a series of elliptical annular apertures with constant width (equal to $1/4$ of the beam's major axis) and increasing values of the inner semi-major axis to cover the whole extension of the disk, from $\Rin'$ to $\Rout'$. The eccentricity of the annulus is given by the sine of the disk inclination, measured with respect to the plane of the sky $51.7^\circ$.

To fit the main characteristics of the ALMA observation, the initial surface density profile had to be modified. This modification was done through the following iterative procedure \rt{which has been used in previous works (e.g. \citealt{Pinte2016,Muro2018,Rab2020})}. The density profile was multiplied point-by-point by the ratio of the observed to modelled azimuthally averaged radial brightness profiles at each radial point. With the modified profile, we repeat the same procedure: run the Monte Carlo simulation, ray-trace the system and convolve the resulting image, extract a new azimuthally averaged radial brightness profile and compare with the observed one. After five iterations, our synthesised radial profile converged to a good extent towards the observed one and the iterations were stopped. The iteratively modified surface density is shown in Fig. \ref{fig:density_profiles} together with the profile reported by \cite{Keppler2018} for comparison.

\begin{figure}[h!]
\centering
\includegraphics[width=1.0\linewidth]{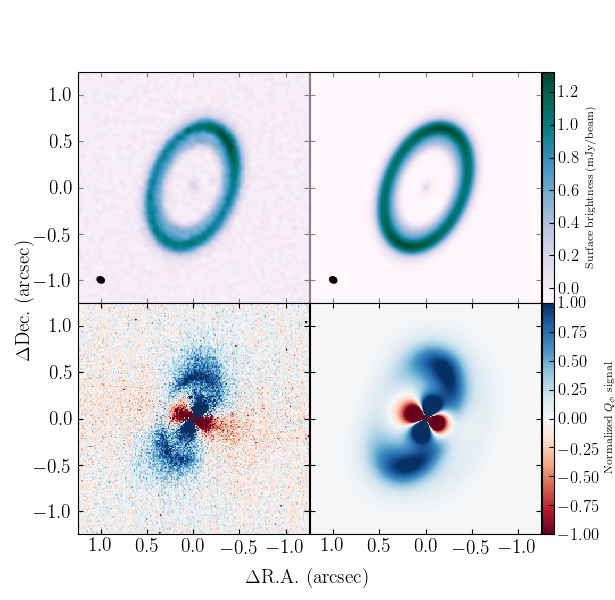}
\caption{\small Comparison between the ray-traced images retrieved from our model (right column) and the ALMA-VLT/SPHERE observations (left column). \textit{Top panels:} thermal emission at $\lambda=855 \, \microns$. \textit{Bottom panels:}  $\Qphi$ component of the polarised and scattered emission at $\lambda=1.25 \, \microns$. North is up, and east is to the left.}
\label{fig:Qphi_alma}
\end{figure}

The top panels in Fig. \ref{fig:Qphi_alma} show a comparison between the observed image and that obtained with our model. Using Eq.(\ref{eq:mdust}) we derive a dust mass of $4.4\times 10^{-5} \, M_\odot$ in the protoplanetary disk, which is slightly higher than the value found by \cite{Keppler2018} via modelling of the near-infrared emission

\rt{The ALMA observation shows an asymmetry in the azimuthal distribution of the brightness in the outer disk which is especially visible along the major axis. The existence of this feature was pointed out by \cite{Long2018}, confirmed by \cite{Keppler2019} and recently by \cite{Benisty2021} with observations at high angular resolution. To quantify this asymmetry, we measured the flux density as a function of the position angle inside a 2 au wide annulus centred at 75 au. We found the ratio of the maximum to minimum values to be 1.5 in the observation but only 1.2 in the simulated image. Our model does not aim to reproduce the observed asymmetry in the outer disk and any asymmetry is smeared out during the process we implemented to find the modified density profile. However, given the quality of the fit that our model provides to the photometric points at (sub-)millimetre wavelengths, and our focus to investigate the CPD, we did not modify the density profile any further and leave the detailed analysis of the asymmetry as a goal for a future study. It is also important to note from Fig. \ref{fig:density_profiles} that we restricted the column density to a minimum of $10^{-5} \, \mathrm{g}\, \mathrm{cm}^{-2}$. This value is in essence determined by the sensitivity of the ALMA observation.}

\rt{Finally, our approach to simulate the ALMA observation by convolving the ray-traced image with a Gaussian beam misses complex details that might arise in the reduction of interferometric data, such as spatial filtering and the presence of correlated noise in the final image. According to the array configuration used to image the source, the maximum recoverable scale is of the order of $1000$ au \citep{Isella2019}. Therefore, this does not affect the analysis of compact emission sources, such as the inner disk or the CPD, given the small angular scales covered by them. On the other hand, we show in Sect. \ref{sec:cpd_environment} that even the fingerprint of the lowest mass CPD considered in this study can be marginally detected above the rms noise level of the observation. The inclusion of these effects, although important for a detailed comparison with the observations, will not modify the fundamental conclusions of our work given the angular resolution and sensitivity of the analysed data set.}

\subsection{Modelling the polarised and scattered light emission}
\label{sect:Qphi_modeling}
Using the results obtained after modelling the ALMA image, we performed an additional ray-tracing at $1.25 \, \microns$ leading to the simulated $Q$ and $U$ Stokes frames for linear polarisation. As the polarised intensity, given by $\mathrm{PI}=\sqrt{Q^2+U^2}$, tends to be a small fraction of the total intensity ($\sim 10\%$), the absolute value of the independent Stokes frames tends to be close to the respective noise level. As a consequence, measurements based on the polarised intensity commonly come with large errors of systematic nature \citep{Schmid2006}. To circumvent this problem, we worked with the polar-coordinates Stokes parameters $\Qphi$ and $\Uphi$ defined as:

\begin{align}
\Qphi&=+Q\cos (2\phi) + U\sin (2\phi)\,, \\
\Uphi&=-Q\sin (2\phi) + U\cos (2\phi)\,;
\end{align}

\noindent where

\begin{equation}
\displaystyle \phi=\arctan \Bigg(\frac{x-x_0}{y-y_0}\Bigg)+\phi_0 \, ,
\end{equation}

\noindent is the angle between the position vector of a pixel with coordinates $(x,y)$ in the $Q$ and $U$ frames, and a unit vector pointing to the north direction, with the origin of both vectors placed at the position of the star $(x_0,y_0)$. The term $\phi_0$ accounts for extra polarisation effects such as those induced by the instruments and was set equal to zero in our models. In this parametrisation, a positive signal in $\Qphi$ accounts for scattered light polarised in the azimuthal direction whereas $\Qphi$ negative implies radial polarisation. The parameter $\Uphi$ contains the light polarised with a direction $\pm 45
^\circ$ with respect to the radially polarised light.

\begin{figure*}[h!]
\centering
\includegraphics[width=1.0\linewidth]{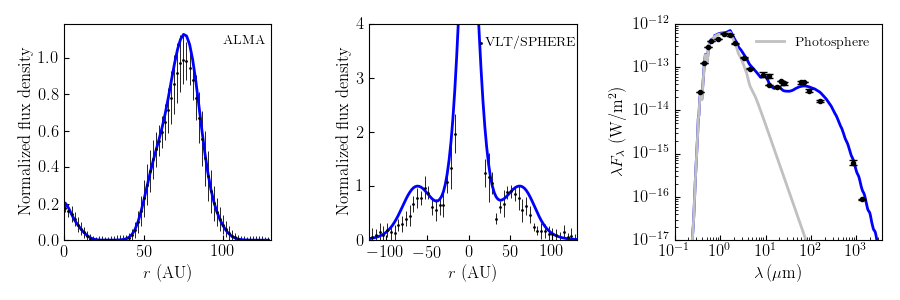}
\caption{\small \textit{Left panel:} Azimuthally averaged radial brightness profile from the ALMA image (black dots) and from our model (blue line). \textit{Middle panel:} Radial cut along the major axis of the $\Qphi$ frames (negative radial direction is along the red-shifted semi-major axis, which is nearly coincident with the south-east direction). \textit{Right panel:} Spectral energy distribution towards PDS 70; black dots correspond to the source's photometry and the blue line is our prediction.}
\label{fig:rt_model}
\end{figure*}

Unlike the disk observed at ALMA wavelengths, the $\Qphi$ frame derived from the observations is clearly non-azimuthally symmetric (see left figure in lower panel of Fig.~\ref{fig:Qphi_alma}). \rt{Instead of trying to fit an azimuthally averaged profile, we aimed to reproduce the $\Qphi$ signal within a conical aperture centred along the major axis which is $\pm 10^\circ$ wide in radial steps of $\sim 3$ au. This corresponds roughly to one-half of the resolution of the IRDIS polarimetric J-band observation (non-coronagraphic case) presented by \cite{Keppler2018}. We did not consider stellar distances smaller than the image resolution of $49 \, \mathrm{mas} \, (\sim 5.6 \, \mathrm{au}$). The uncertainties were computed as the standard deviation of the signal within the correspondent radial bin. The resulting profile retrieved from the observation is shown in black markers in the central panel of Fig. \ref{fig:rt_model}}.

A first comparison with the modelled profile failed at capturing the main features of the $\Qphi$ frame which suggested to modify other parameters of our model. Scattered light is a tracer for the surface layers of protoplanetary disks which are mainly populated by small dust grains whose physical properties determine the polarisation state of the scattered wave. This suggests that a better representation of the data set could be obtained by modifying the physical properties of the dust grains and the geometry of the disk. We motivate the modification of the dust properties upon the assumption that some mechanisms of dust growth and radial drift are taking place in PDS 70.

With \texttt{MCMax3D} we can divide the protoplanetary disk in different zones. Each zone can have a distinct geometry respect to the others, and two zones can overlap to mimic, for example, the presence of an embedded CPD. We divided the protoplanetary disk in two zones: the inner from $0.04$ au to $40$ au and the outer from $40$ au to $130$ au. With these limits, the dust cavity and the inner disk are within the inner zone whereas the main dust ring is in the outer one. Our numerical experiments suggest that this configuration can capture many of the features observed in scattered light. Each zone has its own dust size distribution (see Table~\ref{tab:model_properties}). Our underlying assumption for the dust segregation is motivated by the results of \cite{Bae2019}, where two-dimensional hydrodynamical simulations of the PDS 70 system, led them to demonstrate that large grains are filtrated at the edge of the gap whereas the inner protoplanetary disk is only populated by micron-sized grains.

 With this information at hand, we populated the inner zone with grains of maximum size $100 \, \microns$ whereas the outer zone contains larger grains with a maximum size of $3$ mm. Other bulk properties such as the power law index of the size distribution, the amount of carbon content and the porosity have all equal values in the two zones but slightly different to those used in Sect.\ref{sect:alma_modeling} to fit the ALMA observation. The modification of these properties was also needed to improve the fit of the SED (see Fig. \ref{fig:rt_model} right panel). The turbulence level, controlled by the settling parameter $\alpha$, decreases from the inner to the outer zone. The implemented values are all within the ranges explored by diverse authors for other systems (e.g. \citealt{Muro2018}). Table \ref{tab:model_properties} summarises the whole set of properties of our radiative transfer model. In Fig. \ref{fig:opacities} we show the extinction, absorption and scattering opacities for each zone.

 Our fit to the $\Qphi$ observation can still be improved. This might impose stronger constraints on the population of small size grains and their properties. \rt{We note that we cannot reproduce the azimuthal asymmetry in the brightness distribution of the outer disk, where the NE side appears brighter than the NW side (see Fig. \ref{fig:Qphi_alma} in this work and also Fig. 3 in \citealt{Keppler2018}). This effect might be connected to the scattering properties of the particles populating the surface of the outer disk, since the degree of forward scattering increases when those layers consist of larger grains. Geometrically, the NW side of the disk also represents the near side of the disk and therefore, an enhancement in the number of forward scattered photons due to the larger grains would make this side to appear brighter.} However, since the main goal of the subsequent sections is related to the sub-millimetre regime, we do not aim to pursue a better fit to the infrared data.

\rt{A couple of differences in the assumptions made by \cite{Keppler2018} and this work stand out. Firstly, since our model assumes an inherent inhomogeneity in the radial distribution of grain sizes, our inner and outer disks have different opacities, as can be seen in Fig. \ref{fig:opacities}. At visual wavelengths, the opacity of the inner disk is is a factor of a few larger compared to the outer disk opacity, which is explained by the presence of smaller grains populating the inner disk. This impacts the modelling of structures in the outer disk since we have to compensate for the fewer photons reaching the outer disk by, for example, increasing the outer disk mass. In contrast, the dust opacities in Keppler's model are expected to be the constant throughout the disk and the absolute values are also expected to differ from ours because of the differences in grain composition and size distribution slope\footnote{For further details we refer the reader to \cite{Keppler2018} and references therein and to \cite{Woitke2016} for a description of the DIANA standard opacities used in this work.}. Secondly, the settling of dust in our model is controlled by the Shakura-Sunyaev $\alpha$ parameter through the steady-state solution of the diffusion equation for the dust density (see Sect. \ref{sec:segregation}). This approach differs from \cite{Keppler2018} where the disk was modelled as a two-layer structure with different scale heights to account for the difference in the vertical distribution of large and small grains. Although both approaches provide a good fit to the data, we note that in order to fine-tune the value of the alpha parameter, a close look to the SED is required since this parameter determines the behaviour of the flux in the multicolour and sub-millimetre part of the spectrum. An SED fitting leaves a degeneracy between several parameters, namely the dust mass in the outer disk, the scale height, and the turbulence parameter. That said, we base our subsequent analysis on the parameters listed in Table \ref{tab:model_properties} but we point out that these will not be unique}

%--------------------------------------------------- One column table

\begin{table*}
\caption{Parameters of the radiative transfer model.}
\label{tab:model_properties}
%\resizebox{\columnwidth}{!}{
\centering
\begin{tabular}{llll}
\toprule \toprule
Property & Zone 1 & Zone 2 & Zone 3 (CPD) \\
\midrule
\multicolumn{4}{c}{Physical properties of dust grains}\\
\midrule
Minimum particle size: $\amin \, (\microns)$ & 0.05 & 0.05 & 0.05\\
Maximum particle size: $\amax \, (\microns)$ & 100 & 3000 & 3000\\
Particle size power index: $p$ & 3.9 & 3.9 & 3.5\\
Porosity (\%) & 20 & 20 & 20\\
Amorphous carbon by volume (\%) & 19.2 & 19.2 & 19.2\\
Turbulent settling parameter: $\alpha$ & $2\times 10^{-2}$ & $1\times 10^{-2}$ & $1\times 10^{-4}$\\
\midrule
\multicolumn{4}{c}{Geometrical properties}\\
\midrule
Inner radius: $\Rin$ (au) & 0.04 & 40 & 0.007\\
Outer radius: $\Rout$ (au) & 40 & 130 & 1.2 \\
Flaring exponent: $\beta$ & 1.14 & 1.14 & 1.14\\
Reference radius: $r_0$ (au) & 100 & 100 & 1\\
Scale height at $r_0$: $H_0$ (au) & 10.61 & 10.61 & 0.1\\
\bottomrule
\end{tabular}
%}
\end{table*}

\section{The effect of a planetary mass companion}
\label{sec:including_a_planet}

\subsection{The effect of a planetary companion on the sub-millimetre flux.}
\label{subsection:no_CPD_case}
We explore the fingerprint of a planetary companion assuming no circumplanetary material around it. The implementation of an additional photon source requires defining its luminosity and effective temperature. \cite{Keppler2019} pointed out the existence of a bridge feature located at a position angle of $288
^\circ$ emerging from the outer ring and possibly connecting it with the inner disk. The closeness of this spur to the location of PDS 70 c alongside the inverted P-cygni profile observed in $\Halpha$, may indicate that the planet is still accreting material from the protoplanetary disk. However, the accretion rate $\sim 10^{-2} \, \Mjup\, \mathrm{Myr}^{-1}$ \citep{Haffert2019} makes it unlikely that the planet is going through a runaway phase of growth.

The model for giant planet formation presented in \cite{Ginzburg2019} is applicable to a post-runaway scenario. \cite{Ginzburg2019} have shown that just after the runaway phase, during the runaway cooling, a planet with a two-layered atmosphere (a convective layer surrounded by a radiative one) will contract and increase its mass with nearly similar characteristic times. As both processes are regulated by the cooling efficiency and by the accretion rate respectively, the similarity between these characteristic timescales allows us to formulate the following equivalence:

\begin{equation}
\label{eq:radacc_balance}
4\pi \sigma R_\mathrm{p}^2 T_\mathrm{eff}^4 = \frac{G M_\mathrm{p} \dot{M}_\mathrm{p}}{R_\mathrm{p}},
\end{equation}

\noindent where $M_\mathrm{p}, \, \dot{M}_\mathrm{p}, \, R_\mathrm{p} \, \mathrm{and} \, T_\mathrm{eff}$ represent the mass, mass accretion rate, radius and effective temperature of the planet; $\sigma$ and $G$ are the Stefan-Boltzmann and gravitational constants, respectively. As demonstrated by \cite{Ginzburg2019}, the common assumption of taking $R_\mathrm{p}\approx 2 \, \Rjup$ during the last Myrs of the planet formation (e.g. \citealt{Eisner2015}) is well justified as long as the opacity at the boundary of the atmospheric layers adopts low values of the order of $\sim 10^{-2} \, \mathrm{cm}^2 \, \mathrm{g}^{-1}$. We assumed this and worked with a fixed planetary radius of $R_\mathrm{p}=2\, \Rjup$. Therefore, an estimate for the effective temperature can be derived using Eq. (\ref{eq:radacc_balance}) once a value for the planet mass is assumed.

Several works have constrained the mass of PDS 70 c, finding values ranging from $\sim 0.5$ to $12 \, \Mjup$ (e.g. \citealt{Haffert2019,Mesa2019,Bae2019,Wang2020}) (see Table \ref{tab:pds70c_properties}). We explored a low and a high mass regime given by these values and a third intermediate case with a mass of $5 \, \Mjup$. The spectral energy distribution of the planet is modelled as a blackbody. The location of the planet is taken to match the position angle and the mean orbital radius of the dynamically stable orbital solution for PDS 70 c reported in \cite{Wang2020} and in \cite{Wang2021}, namely, $r_\mathrm{c}=34.3 \, \mathrm{au}$ and $\mathrm{PA}_\mathrm{c}=280.4^\circ$.

We ran one radiative transfer calculation for each value of the companion's mass. We used \rt{$5\times 10^8$} photon packages and the resulting disk structure was ray-traced at $855 \, \microns$. Analysing the temperature structure of the protoplanetary disk, we find that the planet does heat the dust material in its close vicinity, even though the surface density is just $10^{-5} \, \mathrm{g}\, \mathrm{cm}^{-2}$ (see Fig. \ref{fig:density_profiles}). Indeed, a locally enhanced temperature of $\sim 40 \, \mathrm{K}$ is measured inside the computational cell where the planet is located. This is notoriously different to a temperature of $\sim 20$ K that is measured at the diametrically opposed computational cell. These values refer to the midplane temperature \rt{and correspond to the simulation that includes a planetary companion with an effective temperature of $1054$ K \citep{Wang2021}}. The ray-traced images were convolved with the same beam properties described in Section \ref{sect:alma_modeling}. After the convolution, we find that the signal from the planet is smeared out to an extent that it is not possible to differentiate it from the simulation without the planet. Therefore, our simulations suggest that although the luminosity of PDS 70 c can modify the local temperature of the circumstellar material, the observational evidence of such an effect could not be retrieved with ALMA in the continuum. With this result at hand, we proceed to assess if the sub-millimetre signal from the vicinity of the planet can be explained by the presence of circumplanetary material.

\subsection{The effect of a circumplanetary disk around PDS 70 c on the sub-millimetre flux}
\label{sec:CPD_around_c}

\rt{In light of the recent work by \cite{Benisty2021} that constrained the CPD size and the works of \cite{Wang2020,Wang2021} that provide estimates for the temperature and luminosity of PDS 70 c, we investigate the observational fingerprint of a CPD using two different approaches. First, we fix the planet luminosity and temperature and the CPD size to their recent estimates and vary the amount of dust in the CPD to reproduce the observation. In the second approach, we fix the CPD mass found in the previous step and investigate the impact of having a less luminous planet. We defer this second part to the discussion in Sect. \ref{sec:cpd_environment}.}

\subsubsection{Constraining the dust content in the CPD}
\label{sec:constrain_mass_cpd}
\begin{figure*}[h!]
\centering
\includegraphics[width=1.0\linewidth]{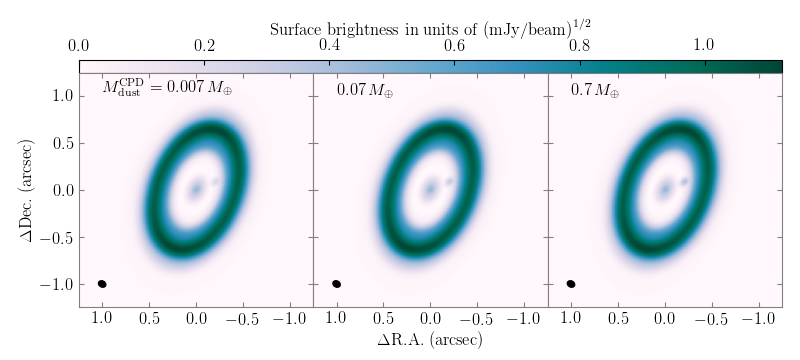}
\caption{\small Synthetic images including a planetary companion surrounded by a CPD. The planet was placed at a distance of $34.3$ au with a position angle of $280.4^\circ$ measured east from north. The three images have similar pixel scales ($4$ mas/px) and were convolved with the same beam properties of the ALMA observation (see Sect. \ref{sect:alma_modeling}). \rt{The colour scale is set equal to the square root of the surface density to boost the signal from the CPD.} North is up, east is to the left.}
\label{fig:CPD_submm}
\end{figure*}

We modelled the CPD as a flared disk centred at the location of the planet. \rt{The outer radius of the CPD was set to $1.2 \, \mathrm{au}$ according to the upper limit set by \cite{Benisty2021}. Assuming this value is equal to the truncation radius of the CPD, which has been shown to scale as one-third of the planet Hill's radius \footnote{For a planet with mass $M_{\mathrm{p}}$ and semi-major axis $a_{\mathrm{p}}$ around a star of mass $M$, the Hill radius is defined as: $\Rhill=a_{\mathrm{p}}\big(M_\mathrm{p}/3M\big)^{1/3}$, with origin at the planet.}  (e.g. \citealt{Quillen1998}) we derive a rough estimate for the planet mass of $2.8 \, \Mjup$.} We set the inner radius of the CPD to $0.007$ au\footnote{\rt{We note that for a $2\, \Rjup$ planet with a luminosity of $5\times 10^{-5} \, \Lsun$ accreting material at $10^{-8} \, \Mjup\, \mathrm{yr}^{-1}$, the dust-sublimation radius where the temperature reaches $1500 \, \mathrm{K}$, lies inside the physical radius of the planet $(R_\mathrm{p} \sim 0.001 \, \mathrm{au})$. Therefore, a CPD inner radius of $0.007 \, \mathrm{au}$ satisfies that the material in the disk is not in direct contact with the surface of the planet and also at an adequate temperature to avoid dust sublimation.}} and we populated it with dust grains ranging from $0.05\, \microns$ to $3\, \mathrm{mm}$ in size with the same content of amorphous carbon and porosity as the grains in the protoplanetary disk. Opacities are again computed using the DIANA standard (see Fig. \ref{fig:opacities}). The rest of the physical and geometrical properties for the CPD are listed in Table \ref{tab:model_properties}. A simulation with $5\times 10^8$ photon packages was ran for three different CPD masses, namely, $0.007 \, \Mearth,\, 0.07 \, \Mearth\, \mathrm{and}\, 0.7 \, \Mearth$. The results of these simulations are shown in Fig. \ref{fig:CPD_submm} and discussed in detail in Sect. \ref{sec:cpd_environment}. Unlike the simulations presented in Sect. \ref{subsection:no_CPD_case} where only the planet was included, it is now evident how the CPD modifies the ray-traced image. The photons emitted by the planet are absorbed and reprocessed by the dust grains in the CPD and their subsequent re-emission boosts the sub-millimetre flux.

\rt{Numerical simulations show that the depletion timescales for the dust component in CPDs are smaller than the respective counterpart in protoplanetary disks. For example, \cite{Rab2019} show, using dust evolution models applied to CPDs around wide-orbit companions, that the dust-to-gas ratio can decrease from $10^{-2}$ to $10^{-4}$ in just 1 Myr. This fast depletion timescale can be mitigated by invoking the presence of dust traps such as those proposed by \cite{Drazkowska2018}. They demonstrated that the advection of gas directed outwards can create an efficient dust trap allowing dust grains to accumulate and grow. In order to constrain the gas-to-dust ratio in CPDs further observations at high angular resolution both in the continuum and in emission lines are required. Therefore, in the absence of observational constraints, we work with a canonical value of $100$ for the gas-to-dust ratio throughout the CPD.}

To try and interpret the result shown in Fig. \ref{fig:CPD_submm} a bit further, at least for the low mass regime where some regions of the CPD are expected to be optically thin, we note that if the CPD is located at a distance $d$ from an observer on Earth with an inclination $i$ with respect to the plane of the sky, then the power at frequency $\nu$ received per unit area is \citep{W2015}:

\begin{equation}
\label{eq:CPD_flux_full}
    \displaystyle \nu F_{\nu}=\frac{2\pi \cos i}{d^2} \int\limits_{\rin}^{\rout}r\nu B_{\nu}\big(T(r)\big)\Bigg[1-\exp{\bigg(-\frac{\tau_{\nu}(r)}{\cos i}}\bigg)\Bigg] \dd r,
\end{equation}

\noindent where the integration limits are determined by the CPD size, $T$ is the dust temperature at a planetocentric position $r$, and $\tau_{\nu}$ is the position-dependent vertical optical depth: $\tau_{\nu}(r)=\kabs \Sigma(r)$. Using the Rayleigh-Jeans formula, Eq. (\ref{eq:CPD_flux_full}) becomes\footnote{$\displaystyle \Tave \equiv \frac{\int r \Sigma(r) T(r) \dd r}{\int r \Sigma(r) \dd r}$}:

\begin{equation}
\label{eq:CPD_flux}
    \displaystyle \nu F_{\nu} \approx \frac{2\nu^3 \kb}{(d c)^2} \kabs \Tave \Mdust,
\end{equation}

 \noindent with $c$, the speed of light. It is clear then that the flux \rt{from an optically thin CPD} scales linearly with its dust content. However, due to the relatively high values of the surface density, especially in the inner regions of the CPD (see Sect. \ref{sec:temp_cpd}), we expect optically thick emission from the CPD. \rt{In fact, analysing the density profile of the disk with a value for the absorption opacity of $8 \, \mathrm{cm}^2\, \mathrm{g}^{-1}$ (see Fig. \ref{fig:opacities}), we found that only the $0.007 \, \Mearth$ disk has an optical depth lower than 1 at $855 \, \microns$ for $r>0.01 \, \mathrm{au}$. The two more massive disks are entirely optically thick. A full numerical radiative transfer treatment for the study of these compact structures seems thus justified.}

\begin{figure*}[h!]
\begin{subfigure}{0.5\textwidth}
  \centering
  \includegraphics[width=1.0\linewidth]{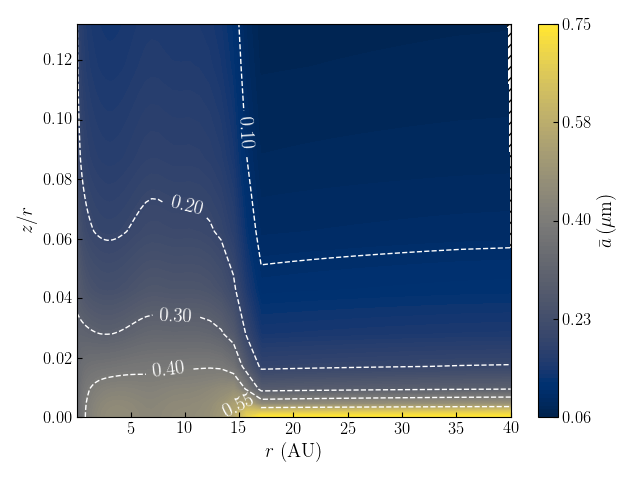}
%\caption{1a}
  \label{fig:sfig1}
\end{subfigure}%
\begin{subfigure}{.5\textwidth}
  \centering
  \includegraphics[width=1.0\linewidth]{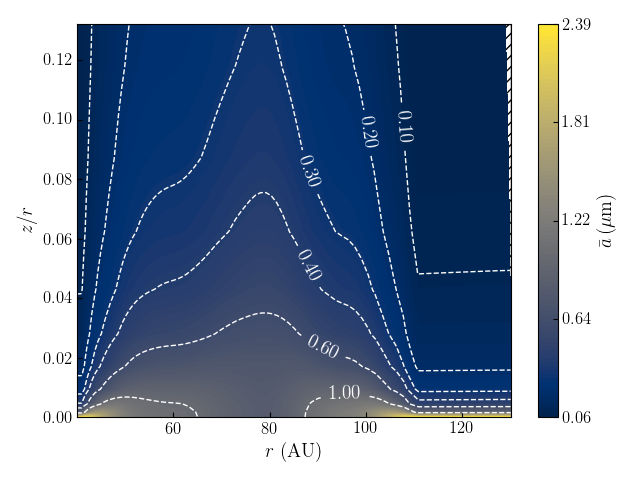}
  %\caption{1b}
  \label{fig:sfig2}
\end{subfigure}
\caption{\textit{Left:} Cut in the vertical distribution of mean particle sizes in the inner zone of the protoplanetary disk. Labels along the dashed contours show the mean particle size in $\microns$. \textit{Right:} Same as the firgure on the left, but for the outer zone.}
\label{fig:vertical_mix}
\end{figure*}

\begin{table}
\caption{Masses, luminosities, and effective temperatures reported in the literature or derived through Eq.(\ref{eq:radacc_balance}) for PDS 70 c.}
\label{tab:pds70c_properties}
%\resizebox{\columnwidth}{!}{
\centering
\begin{tabular}{llll}
\toprule \toprule
Reference & $M\, (\Mjup)$ & $L\, (\Lsun)$ & $T$ (K)\\
\midrule
\multicolumn{4}{c}{PDS 70 c}\\
\midrule
\cite{Wang2021} &  & $4.53\times 10^{-5}$\tablefootmark{a} & 1054.0\\
\cite{Wang2020} & $1.0-3.0$ & $3.60\times 10^{-5}$ & 995.0\\
\cite{Mesa2019} & $0.5-4.0$ & $2.49\times 10^{-5}$\tablefootmark{a} & 900.0\\
\cite{Bae2019} & $10.0$ & $1.39\times 10^{-5}$\tablefootmark{b} & 778.0\tablefootmark{b}\\
\cite{Haffert2019} & $4.0-12.0$ & $1.60\times 10^{-4}$ & 1432.0\tablefootmark{c}\\
\midrule
\multicolumn{4}{c}{Estimates from the \cite{Ginzburg2019} model}\\
\midrule
Low mass case & $0.5$ & $7.0\times 10^{-7}$ & 368.0\\
Intermediate case & $5.0$ & $7.0\times 10^{-6}$ & 654.0\\
High mass case & $12.0$ & $1.6\times 10^{-5}$ & 814.0\\
\bottomrule
\end{tabular}
%}
\tablefoot{\tablefoottext{a}{Derived assuming a radius of $2\, \Rjup$.} \tablefoottext{b}{Values correspond to the model by \cite{Ginzburg2019} for a mass accretion rate of $10^{-8} \, \Mjup \mathrm{yr}^{-1}$.}
\tablefoottext{c}{Value derived assuming $L$ equal to the accretion luminosity}}
\end{table}

\section{Results and discussion}
\label{sec:discusion}
\subsection{Particle size segregation}
\label{sec:segregation}

Our modelling requires to segregate dust grains by size since the initial implementation of a radially homogeneous distribution of small and large grains did not provide a good fit to the scattered light observation and multiple photometric points were too far from the predicted SED. We do not attempt to have a consistent dust evolutionary model aimed to explain the dust grain segregation but we note that such behaviour has been implemented also in other works to explain the observations of protoplanetary disks (e.g. \citealt{Muro2018}; \citealt{Villenave2019}).

In general, small dust particles tend to be well coupled to the gas component which is moving at sub-Keplerian velocities as a result of radial pressure gradients in the disk. Although the motion of larger grains is not directly affected by the pressure gradient, the interaction with smaller grains and gas particles will make the former lose angular momentum and migrate inwards \citep{Birnstiel2010}. However, this migration can be halted via several mechanisms. Some of those invoke the presence of dead zones or giant planets. Dead zones are weakly ionised regions in a protoplanetary disk characterised by the suppression of gas flow \citep{Pinilla2016}. The accumulation of gas at the outer rim of the dead zone leads to an increment in the pressure facilitating the trapping of large grains \citep{Villenave2019}. On the other hand, a giant planet with a mass larger than $1\, \Mjup$ can open a gap in the gas distribution leading again to the enhancement of the pressure at the outer border of the gap. In particular, the scenario involving a giant planet may apply to PDS 70 where at least two planets have been detected. In the case of a two-planet system, we note that \cite{Pinilla2015} carried out hydrodynamical and dust evolution simulations and find that, depending on the disk viscosity and fragmentation velocity of the grains, the presence of a second planet can modulate another overdensity region independent of the one created by the first planet. Regardless the operating scenario, only the smaller particles can penetrate the pressure wall and migrate inwards while the larger ones remain trapped \citep{Fouchet2007}.

\subsection{Vertical distribution of dust grains}
\label{sec:vertical_segregation}
In our models, the vertical stratification of grains is implemented following the prescription of \cite{Dubrulle1995} where the settling is the result of a diffusion process mediated by the gas turbulence and the gravitational effect of the central body. The strength of the turbulence is proportional to the turbulent viscosity and can be parameterised by the dimensionless $\alpha$ parameter introduced by \cite{Shakura1973}. Prior to the formal Monte Carlo iteration, the steady-state solution of the diffusion equation is computed to find the dust density as a function of the vertical coordinate for each bin of the size distribution.

\rt{The steady-state solution to the diffusion equation can be written in terms of the scale height $H_\mathrm{d}(r,a)$ of a dust grain with size $a$. This expression is simply \citep{Dubrulle1995,Woitke2016}:
\begin{equation}
\bigg(\frac{H_\mathrm{d}(r,a)}{H(r)}\bigg)^2=\frac{\alpha (1+\gamma_0)^{-1/2}}{a \rho_\mathrm{gr}} \rho_\mathrm{mid}(r) H(r),
\end{equation}}
\noindent \rt{where, $\gamma_0\approx 2$ for compressible turbulence, $\rho_\mathrm{gr}$ is the mass density of a dust grain ($\rho_\mathrm{gr}=2.18 \, \mathrm{g}\,\mathrm{cm}^{-3}$ in our simulations) and $\rho_\mathrm{mid}$ is the gas density in the midplane (see Eq. \ref{eq:gas_density}). Therefore, for each cell in the disk we identify a particle size distribution of the form $f(a,\vec{r}) \propto \mathrm{exp}(-z^2)/[2H_\mathrm{d}(r,a)^2]$, where the vector $\vec{r}$ denotes the position of the cell in the three-dimensional grid and $z=r\cos \theta$ is the height above the midplane for a point with co-latitude $\theta$.}

To visualise the effect of the vertical segregation we first define the mean grain size via the third moment of the size distribution:

\begin{equation}
\bar{a}(\vec{r}) \equiv \amean (\vec{r})=\Bigg[\frac{1}{\bar{n}}\int\limits_{\amin}^{\amax} f(a,\vec{r})a^3 \dd a \Bigg]^{1/3},
\end{equation}

\noindent where $\bar{n}$ is the total number density of representative equal-sized particles as used in \cite{Vasyunin2011}. The resulting structure is computed for both zones in the protoplanetary disk and shown in Fig. \ref{fig:vertical_mix}. The shape of the contours is clearly related to the surface density profile which in turn modulates the shape of the gas density.

Inside the gap, the level of settling is enhanced since the surface density drops to $10^{-5}\, \mathrm{g}\, \mathrm{cm}^{-2}$ (see Fig. \ref{fig:density_profiles}); the larger particles in the inner zone are settled to the midplane whereas for $z$ values close to one scale height, the smaller particles prevail. On the other hand, the right panel of Fig. \ref{fig:vertical_mix} shows how particles in the outer zone are vertically segregated. The main dust ring shows a low degree of settling, even though the value of $\alpha$ is \rt{slightly lower} than in the inner zone (see Table \ref{tab:model_properties}). This is again related to the high value of the gas density around $80$ au. In Fig. \ref{fig:vertical_mix}, it is observed an apparent lack of grains with size close to the $\amax$ value of the respective zone. This can be understood considering the following arguments. First, this has to do with the relatively high value of the exponent of the size distribution used in the model (see Eq. \ref{eq:size_distribution}). Such a high exponent increases the steepness of the distribution function which in turn amplifies the statistical weight of the smaller grains. Second, it is also a numerical effect due to our definition of the mean particle size which is taken to be proportional to the cubic root of the third moment of the distribution. We note that other definitions are also suitable and they can make explicit the presence of larger particles in Fig. \ref{fig:vertical_mix} (e.g. \citealt{Vasyunin2011,Facchini2017}). However, the conclusions about the vertical stratification will remain unaltered regardless the formalism used to define the mean particle size.

\subsection{Constraining the dust environment around PDS 70 c}
\label{sec:cpd_environment}
\begin{figure}[h!]
\centering
\includegraphics[width=1.0\linewidth]{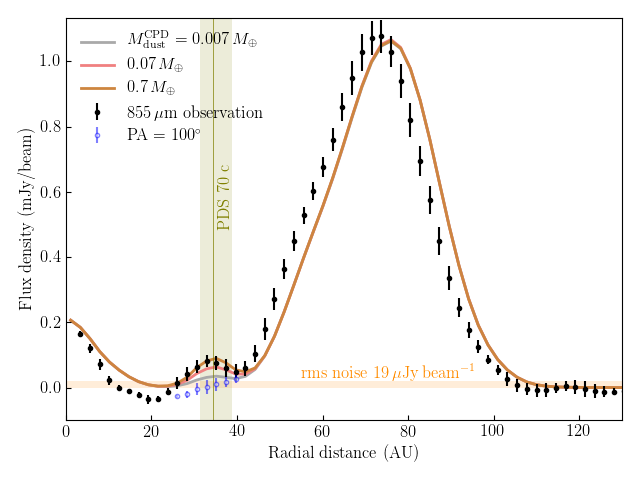}
\caption{\small Radial cuts extracted from Fig. \ref{fig:CPD_submm}. The conical aperture was centred at a position angle of $280.4^\circ$ in line with the astrometric position of PDS 70 c. The aperture has a width of $\pm 10^\circ$ around the mean position and the radial step size is $2$ au.}
\label{fig:cpd_three_cases}
\end{figure}

A visual inspection of the ALMA data set (Fig. \ref{fig:Qphi_alma}) suggests the presence of a dim spot nearly matching the location of the dynamically stable orbital solution for PDS 70 c in \cite{Wang2021}. \rt{Such a feature} can be retrieved from our simulations when we include a CPD that is massive enough, as shown in Fig. \ref{fig:CPD_submm}. Therefore, we use our models to measure the brightness profile in the vicinity of PDS 70 c and aim to draw some conclusions about the expected mass ranges of the CPD. \rt{We extracted the radial brightness profile using a conical aperture with $\pm 10^\circ$ width around the position angle of PDS 70 c} ($\mathrm{PA}=280.4^\circ$ measured east-from-north) with radial bins equally spaced by $0.017$ arcsec ($\sim 2$ au), equivalent to one quarter of the ALMA beam's major axis. This procedure was performed with the GoFish package \citep{Teague2019}. Figure \ref{fig:cpd_three_cases} shows the cuts obtained from the observation \rt{compared to the three CPD mass cases simulated in this work.}

\rt{Although the width and height of the modelled curves fit the observation, all the curves show an offset of $\sim 2$ au with respect to the observation. It might be that our fitting procedure based on extracting azimuthal averages is not well suited to reproduce the observed flux along individual directions especially when they present evident asymmetries. We do not attempt to dive into this issue. Despite the mismatch, our results can be used to explore the dust content in the CPD. The blue points in the interval $ 26 \lesssim r \lesssim 40$ au were extracted from the side of the disk opposite to the direction of PDS 70 c (i.e. $\mathrm{PA} \approx 100^\circ$) where there is no evidence of emission other than the background continuum. Due to the level of dust depletion in the cavity, this background emission and its uncertainty allow us to conclude that a $0.007 \, \Mearth$ CPD would be marginally detectable with the sensitivity of our data set. Figure \ref{fig:cpd_three_cases} also depicts the $19 \, \mu\mathrm{Jy}\, \mathrm{beam}^{-1}$ rms noise level of the observation as reported by \cite{Isella2019}.} The photometric curves in Fig. \ref{fig:cpd_three_cases} indicate that the model with a $0.7\, \Mearth$ CPD predicts a flux density that agrees with the observed values within the error bars.

\rt{Our estimate for the mass of PDS 70 c of $\sim 2.8 \, \Mjup$ is very close to the upper value estimated by \cite{Wang2020} via SED fitting (see Table \ref{tab:pds70c_properties}). This is remarkable considering that we used a very simple argument, namely, that the CPD radius equals one-third of the Hill's radius. However, as noted by \cite{Wang2020}, the extent of their wavelength coverage opens the question whether or not the luminosity they derived properly approximates to the bolometric luminosity needed to estimate the mass of the planet with the evolutionary model of \cite{Ginzburg2019}. Part of the bolometric luminosity could be captured by the CPD and then radiated away at wavelengths longer than $\sim 4\, \microns$, which is the limit of the wavelength coverage of Wang's data set. If this is the case, an infrared SED fitting with a limited wavelength coverage will underestimate the total luminosity because the SED is expected to peak at longer wavelengths \citep{Zhu2015,Szulagyi2019}. Consequently, the planet mass will be underestimated. In our models, the assumed mass of the planet affects the dust settling within the CPD and this could alter our predictions for the sub-millimetre flux. Additional observations at longer wavelengths are needed to obtain more robust estimates to the mass of PDS 70 c via SED fitting.}

\begin{figure}[h!]
\centering
\includegraphics[width=1.0\linewidth]{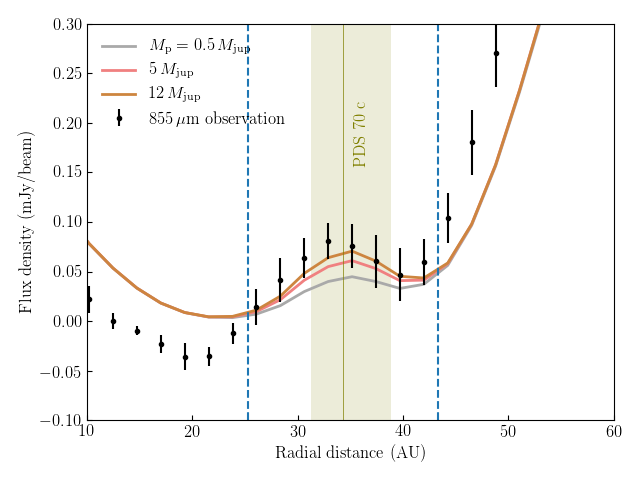}
\caption{\small Effect on the CPD flux due to the adopted planet mass and luminosity. In this experiment, our reference CPD mass was fixed to $0.7 \, \Mearth$. Vertical dashed lines represents the limits of the circular aperture used to measure the flux emitted by the CPD.}
\label{fig:cpd_mass_dependence}
\end{figure}

\rt{The amplitude of the crest at 34 au in Fig. \ref{fig:cpd_three_cases} depends not only on the CPD mass but also on the planet properties, particularly on its luminosity and mass. In order to test how sensitive our results are to these variables, we abandon the estimates provided in \cite{Wang2021} and use instead the theoretically expected values from the \cite{Ginzburg2019} model for the low, intermediate and high planet mass cases (see Table \ref{tab:pds70c_properties}). In all cases we fixed the size of the CPD to $1.2$ au in accordance to the most recent observations \citep{Benisty2021}. The mass of the CPD was fixed to $0.7 \, \Mearth$ and the results are shown in Fig. \ref{fig:cpd_mass_dependence}. Evidently, the assumed properties for the planet impact the flux emitted by the optically thick CPD. Since the dust mass was kept fixed and even the highest planetary luminosity is still lower than Wang's luminosity, it is not a surprise that the observations tend to be better explained by high mass planets. Given the results presented in this subsection, we claim that it is not possible to constrain individually the mass of the planet and the dust budget in the CPD.}

We performed aperture photometry in the ray-traced images using a circular aperture $9$ au in radius around the planet's mean location. When the CPD mass is $0.07\, \Mearth$, the measured flux is $64\%$ of the $0.7\, \Mearth$ case and when the CPD mass is $0.007\, \Mearth$ the flux represents just $\sim 37\%$. This indicates that at least one of the assumptions made going from Eq. (\ref{eq:CPD_flux_full}) to Eq. (\ref{eq:CPD_flux}) (which predicts a linear scaling) breaks and we reinforce the idea that the CPD cannot be treated as optically thin.

\rt{We can also compare the values of the peak intensities between our model and the ALMA observation. \cite{Isella2019} reported a CPD flux peak value of $106\pm 19 \, \mu\mathrm{Jy} \, \mathrm{beam}^{-1}$ whereas our model predicts: $150$, $106$ and $58\, \mu\mathrm{Jy}\, \mathrm{beam}^{-1}$ for the $0.7$, $0.07$ and $0.007\, \Mearth$ CPD mass cases, respectively. Similar to the brightness profiles in Fig. \ref{fig:cpd_three_cases}, this result also suggests that the mass of the CPD is constrained to the range $0.07-0.7 \, \Mearth$.}

\subsection{Comparison with recent works}
\rt{\cite{Benisty2021} observed the continuum emission at $\lambda=855 \, \microns$ with an angular resolution of $\sim 20\, \mathrm{mas}$. The observation clearly shows a localised signal in the expected location of PDS 70 c and it is
detached from the disk although still unresolved. The authors interpret this sub-millimetre emission as originating from dusty material in a CPD.}

\rt{They detected emission from a resolved inner disk and argue that the presence of the planets can produce an effective mechanism for dust trapping, so only the small dust grains can flow towards the inner disk. This agrees with our findings, where the particles in the inner disk have sizes from $0.05$ to $100$ $\microns$. Our model for the protoplanetary disk predicts a flux from the inner disk equal to $0.76 \, \mathrm{mJy}$, which is very close to the observed mean value of $0.85 \, \mathrm{mJy}$. We estimated the dust mass in the inner disk to be $0.02 \, \Mearth$. This value is low in comparison to the $0.08 \, \Mearth$ lower limit estimated in \cite{Benisty2021}. This difference could be explained by the steepness of our size distribution ($p=-3.9$). This causes the abundance of large grains in the inner zone to be lower than in Benisty's work ($p=-3.5$). Combining the result in Fig. \ref{fig:density_profiles} with an absorption opacity of $8\, \mathrm{cm}^2\, \mathrm{g}^{-1}$ suggests that the inner disk is optically thin at $\lambda=855\, \microns$.}

\rt{Working under the assumption of an optically thin CPD populated with submicron and centimetre-sized grains, \cite{Benisty2021} estimated the dust mass in the CPD as a function of the maximum grain size. In particular, when the dust size exponent is assumed to be $-3.5$, they found a mass of $\sim 0.009 \, \Mearth$. Our $0.007\, \Mearth$ CPD case does not provide a satisfactory fit to the observation. As our more massive CPDs become optically thick, we only comment on the comparison of these two low-mass cases. First of all, the semi-analytical estimate of \cite{Benisty2021} only depends on the absorption opacity. However, at mm wavelengths and assuming a size distribution with maximum size above $100 \, \microns$, the scattering opacity exceeds the absorption opacity \citep{Birnstiel2018}. A similar behaviour is observed in the DIANA opacities (see Fig. \ref{fig:opacities} for the CPD zone). Therefore, the inclusion of scattering effects is expected to alter the emitted flux even in optically thin systems \citep{Zhu2019}. Our model takes into account those non-isotropically scattered photons that will not contribute to the dust heating. As a result, the same amount of dust will emit a lower millimetre flux when scattering is considered compared to the case when scattering is neglected.}

\rt{Another point relates to the assumed CPD temperature. \cite{Benisty2021} estimated the dust temperature from the combined effect of several heating sources: the viscous accretion and the planet and stellar irradiation at the location of PDS 70 c. With this assumptions, they found a temperature of $\sim 26$ K at 1 au. In contrast, our simulation predicts a temperature of $10$ K at the same distance. Since the dust mass is inversely proportional to the temperature in an optically thin disk (see Eq. \ref{eq:CPD_flux}), our lower estimate for the temperature requires a higher dust mass to reproduce the same flux.}

\rt{Finally, we note that although most of the emission from the $0.007 \, \Mearth$ CPD is expected to be optically thin, our model suggests that the innermost region could still be optically thick because of the dust density decaying exponentially with distance. This, of course, is just an assumption since spatially resolved observations of CPDs are still not available. However, it raises the question whether or not the heating of those grains in the outskirts is due to photons from the planet's photosphere travelling freely through the medium. Infrared photons re-emitted by the optically thick layers might have an important role setting the equilibrium temperature of the outer CPD. We note that the potential effects of optically thick emission are accounted in our model.}

\subsection{The thermal environment of the CPD}
\label{sec:temp_cpd}
Since the CPD was modelled as an additional zone placed on top of the protoplanetary disk, the thermal structure was self-consistently calculated during the \rt{same Monte-Carlo run that allowed us to find the temperature in the protoplanetary disk}. In our simulations, the dust temperature in the CPD is set by the stellar photons reprocessed by the protoplanetary disk and by the planet luminosity\footnote{We note that our simulations did not include the heating effect of the accretion luminosity which may alter the temperature profile of the CPD. Given the rate of mass accretion for PDS 70 c ($\sim 10^{-8} \, \Mjup\, \mathrm{yr}^{-1}$ \citealt{Haffert2019}), the contribution of the accretion luminosity to the midplane temperature at the outer edge of the CPD would be nearly $3$ K, according to the models of \cite{Dalessio1998} and \cite{Isella2014}.}, \rt{after a prescription for the vertical structure of the CPD has been provided (see Table \ref{tab:model_properties})}. We aim to disentangle the contribution to the total CPD temperature of both sources by switching on and off the effect of the planet luminosity in the simulation.

\begin{figure}[h!]
\centering
\includegraphics[width=1.0\linewidth]{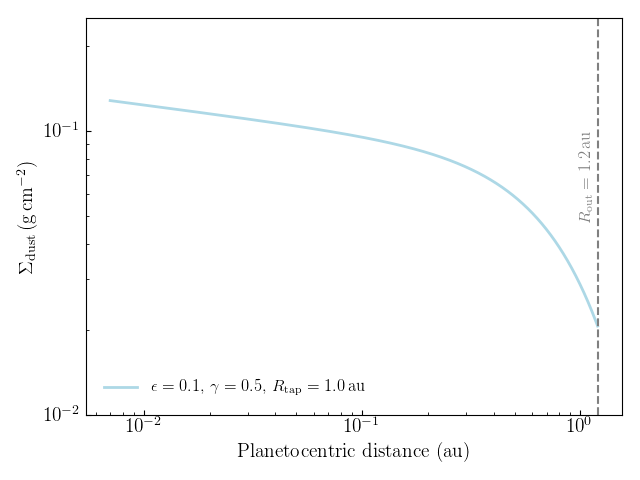}
\caption{\small \rt{Surface density profile for an optically thin CPD at sub-millimetre wavelengths of $1.2$ au in size obtained from Eq. \ref{eq:surface_density} using the parameters shown in the lower-left corner. The dust mass is $0.007\, \Mearth$.}}
\label{fig:cpd_surf_density}
\end{figure}

\rt{We use the reference model for the protoplanetary disk described in Sect. \ref{sec:rt_model} and include a $2.8\, \Mjup$ planet with the same estimates for the luminosity and effective temperature as given by \cite{Wang2021}. We also include an optically thin CPD with a dust mass of $0.007 \, \Mearth$ and a size of $1.2\, \mathrm{au}$. Considering an absorption opacity of $8.5 \, \mathrm{cm}^2\, \mathrm{g}^{-1}$ and the density profile of Fig. \ref{fig:cpd_surf_density}, the CPD will be optically thin at sub-millimetre wavelengths down to $0.01\, \mathrm{au}$; this ensures that a high enough number of photon packages is able to reach the disk's midplane and a reliable dust temperature estimate can be then retrieved.}

\begin{figure}[h!]
\centering
\includegraphics[width=1.0\linewidth]{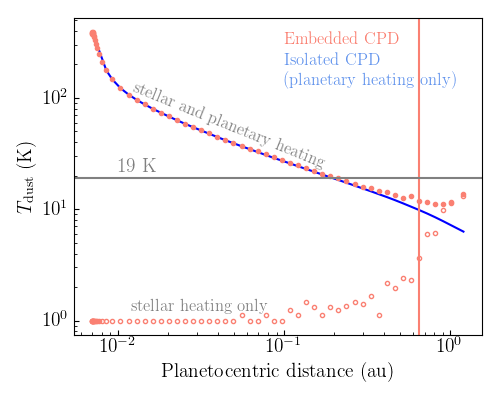}
\caption{\small \rt{Azimuthally averaged midplane temperature for an optically thin CPD at sub-millimetre wavelengths around a $2.8\, \Mjup$ planet. Filled circles indicate the temperature retrieved from the simulation that includes to the planet and to the star as heating sources; open circles indicate the temperature measured when the planet was not included. Vertical line indicates the planetocentric distance beyond which the star alone provides more than $50\%$ of the total CPD heating. The blue curve shows the case of the isolated planet+CPD system where the heating from the circumstellar material is not modelled. The horizontal line marks the characteristic temperature of the protoplanetary disk at the location of the planet.}}
\label{fig:cpd_temperature}
\end{figure}

\rt{The whole PDS 70 system was simulated with and without the planet and the azimuthally averaged midplane temperature was extracted afterwards for both cases. Results are shown with red points in Fig. \ref{fig:cpd_temperature}. The filled circles represent the temperature obtained when the planet was included and the open circles when the planet was removed from the simulation. To quantify the relative contribution of the star to the CPD heating, we computed the ratio between the simulations and identified the distance where it becomes larger than $0.5$. Beyond this location, the heating due to stellar irradiation (direct or scattered by the protoplanetary disk) contributes more than $50\%$ to the total CPD heating. This location is marked with the vertical red line. The blue curve in Fig. \ref{fig:cpd_temperature} is the temperature of an isolated CPD with the same properties but not embedded within a protoplanetary disk so the only heating source is the planet. The horizontal line in Fig. \ref{fig:cpd_temperature} at $19$ K represents the predicted temperature of the protoplanetary disk at the planet location.}

\rt{From Fig. \ref{fig:cpd_temperature}, we can infer that the planet luminosity is the main heating source in the CPD out to $0.6$ au but beyond this radius the temperature is dominated by the stellar heating, either in the form of direct irradiation or in the form of photons re-emitted and scattered by the protoplanetary disk.}

\rt{We conclude that the temperature of dust grains in the midplane of a CPD that is embedded in the dust cavity of PDS 70 at $34$ au is determined by two heating mechanisms which dominate at different spatial locations. Additionally, we can see that the temperature profile is monotonically decreasing when the planet+CPD system is isolated but when it is embedded in the natal protoplanetary disk a reversal of the temperature happens at $\sim 0.8 \, \mathrm{au}$. As a result of this reversal, the predicted temperature at the outer edge of the CPD is $13$ K in the embedded case but only $6$ K in the isolated case.}

\rt{The predicted temperature reversal in the outskirts demonstrates the relevance of a self-consistent radiative transfer modelling of a CPD embedded in the circumstellar environment. This result motivates further studies about the position of icelines in CPDs and their impact on the formation of exomoons.}

\section{Conclusions}
\label{sec:conclusions}
In this work we present a three-dimensional radiative transfer model that jointly explains the observed continuum emission from the PDS 70 system at $855 \, \microns$ and at $1.25 \, \microns$ in the $\Qphi$ component of the polarised and scattered light. To this model, we then added a second source of photons that mimics the planetary mass companion PDS 70 c and the circumplanetary disk (CPD) around it. Here, we summarise our results:

\begin{enumerate}
\item Our simulations show that dust grains are spatially segregated by size in the PDS 70 system. The inner disk is populated by small grains with a \rt{maximum size of $100\,\microns$} and the outer disk is populated by \rt{larger grains with a maximum size of $3$ mm}. The vertical settling is controlled by turbulence parameters equal or lower than $2\times 10^{-2}$. Other bulk parameters such as the carbon content and the porosity are considered to be constant throughout the disk. These properties for the dust grains, alongside a density profile modulated by the substructures observed with ALMA, allow us to reproduce the sub-millimetre and the near-infrared polarised-scattered emission from the PDS 70 system.

\item \rt{The inclusion of a planetary companion of $2.8\, \Mjup$ with values of the effective temperature and luminosity constrained by the work of \cite{Wang2021} (i.e. $T_\mathrm{p}=1054$ K and $L_\mathrm{p}=4.5\times 10^{-5} \, \Lsun$) does not generate any detectable signal at sub-millimetre wavelengths if the model is convolved with a Gaussian beam of size $67 \times 50$ mas. Since the observation shows a clear substructure around the location of the planetary companion PDS 70 c, our models support the presence of an optically thick circumplanetary disk that surrounds the planet.}

\item \rt{Given our assumptions on the planet luminosity and grain properties, and using the most recent estimate for the upper limit for the size of the CPD, we constrained the dust mass in the CPD to the range $0.07-0.7 \, \Mearth$.}

\item For planetocentric distances lower than $0.6$ au, the thermal structure in the midplane of the CPD around PDS 70 c is primarily determined by the planet luminosity. For larger distances, photons emitted by the star and reprocessed by the protoplanetary disks are the dominant heating source.
\end{enumerate}

Although we adopted the premise of a CPD orbiting PDS 70 c, we point out that other interesting scenarios have been proposed, such as the effect of accretion streams and their ability to reproduce some of the observed features without invoking the presence of a CPD (e.g. \citealt{Toci2020}). \rt{However, the recent high resolution continuum observations of PDS 70 \citep{Benisty2021} seem to favour the CPD scenario. Spatially resolved observations of the gas dynamics in the vicinity of PDS 70 c will help to complete the puzzle of this interesting planet forming region.}

\begin{acknowledgements}
We thank the anonymous referee for a constructive report that improved the quality of the paper. This work is partly supported by the Netherlands Research School for Astronomy (NOVA). CHR acknowledges support from the DFG Research Unit "Transition Disks" project number 325594231 (FOR 2634/1 and FOR 2634/2). CHR is grateful for support from the Max Planck Society.
\end{acknowledgements}

\bibliographystyle{aa}
\bibliography{references}

\begin{thebibliography}{56}
\expandafter\ifx\csname natexlab\endcsname\relax\def\natexlab#1{#1}\fi

\bibitem[{Armitage(2010)}]{Armitage2010}
Armitage, P. 2010, Astrophysics of Planet Formation (Cambridge University
  Press)

\bibitem[{{Bae} {et~al.}(2019){Bae}, {Zhu}, {Baruteau}, {Benisty}, {Dullemond},
  {Facchini}, {Isella}, {Keppler}, {P{\'e}rez}, \& {Teague}}]{Bae2019}
{Bae}, J., {Zhu}, Z., {Baruteau}, C., {et~al.} 2019, \apjl, 884, L41

\bibitem[{{Benisty} {et~al.}(2021){Benisty}, {Bae}, {Facchini}, {Keppler},
  {Teague}, {Isella}, {Kurtovic}, {P{\'e}rez}, {Sierra}, {Andrews},
  {Carpenter}, {Czekala}, {Dominik}, {Henning}, {Menard}, {Pinilla}, \&
  {Zurlo}}]{Benisty2021}
{Benisty}, M., {Bae}, J., {Facchini}, S., {et~al.} 2021, \apjl, 916, L2

\bibitem[{{Birnstiel} {et~al.}(2010){Birnstiel}, {Dullemond}, \&
  {Brauer}}]{Birnstiel2010}
{Birnstiel}, T., {Dullemond}, C.~P., \& {Brauer}, F. 2010, \aap, 513, A79

\bibitem[{{Birnstiel} {et~al.}(2018){Birnstiel}, {Dullemond}, {Zhu}, {Andrews},
  {Bai}, {Wilner}, {Carpenter}, {Huang}, {Isella}, {Benisty}, {P{\'e}rez}, \&
  {Zhang}}]{Birnstiel2018}
{Birnstiel}, T., {Dullemond}, C.~P., {Zhu}, Z., {et~al.} 2018, \apjl, 869, L45

\bibitem[{{Canup} \& {Ward}(2002)}]{Canup2002}
{Canup}, R.~M. \& {Ward}, W.~R. 2002, \aj, 124, 3404

\bibitem[{{Carrasco-Gonz{\'a}lez} {et~al.}(2009){Carrasco-Gonz{\'a}lez},
  {Rodr{\'\i}guez}, {Anglada}, \& {Curiel}}]{Carrasco2009}
{Carrasco-Gonz{\'a}lez}, C., {Rodr{\'\i}guez}, L.~F., {Anglada}, G., \&
  {Curiel}, S. 2009, \apjl, 693, L86

\bibitem[{{D'Alessio} {et~al.}(1998){D'Alessio}, {Cant{\"o}}, {Calvet}, \&
  {Lizano}}]{Dalessio1998}
{D'Alessio}, P., {Cant{\"o}}, J., {Calvet}, N., \& {Lizano}, S. 1998, \apj,
  500, 411

\bibitem[{Draine(2010)}]{Draine2010}
Draine, B. 2010, Physics of the Interstellar and Intergalactic Medium,
  Princeton Series in Astrophysics (Princeton University Press)

\bibitem[{{Dr{\k{a}}{\.z}kowska} \& {Szul{\'a}gyi}(2018)}]{Drazkowska2018}
{Dr{\k{a}}{\.z}kowska}, J. \& {Szul{\'a}gyi}, J. 2018, \apj, 866, 142

\bibitem[{{Dubrulle} {et~al.}(1995){Dubrulle}, {Morfill}, \&
  {Sterzik}}]{Dubrulle1995}
{Dubrulle}, B., {Morfill}, G., \& {Sterzik}, M. 1995, \icarus, 114, 237

\bibitem[{Eisner(2015)}]{Eisner2015}
Eisner, J.~A. 2015, The Astrophysical Journal, 803, L4

\bibitem[{{Facchini} {et~al.}(2020){Facchini}, {Benisty}, {Bae}, {Loomis},
  {Perez}, {Ansdell}, {Mayama}, {Pinilla}, {Teague}, {Isella}, \&
  {Mann}}]{Facchini2020}
{Facchini}, S., {Benisty}, M., {Bae}, J., {et~al.} 2020, \aap, 639, A121

\bibitem[{{Facchini} {et~al.}(2017){Facchini}, {Birnstiel}, {Bruderer}, \& {van
  Dishoeck}}]{Facchini2017}
{Facchini}, S., {Birnstiel}, T., {Bruderer}, S., \& {van Dishoeck}, E.~F. 2017,
  \aap, 605, A16

\bibitem[{{Fouchet} {et~al.}(2007){Fouchet}, {Maddison}, {Gonzalez}, \&
  {Murray}}]{Fouchet2007}
{Fouchet}, L., {Maddison}, S.~T., {Gonzalez}, J.~F., \& {Murray}, J.~R. 2007,
  \aap, 474, 1037

\bibitem[{{Gaia Collaboration} {et~al.}(2018){Gaia Collaboration}, {Brown},
  {Vallenari}, {Prusti}, {de Bruijne}, {Babusiaux}, {Bailer-Jones}, {Biermann},
  {Evans}, {Eyer}, {Jansen}, {Jordi}, {Klioner}, {Lammers}, {Lindegren},
  {Luri}, {Mignard}, {Panem}, {Pourbaix}, {Randich}, {Sartoretti}, {Siddiqui},
  {Soubiran}, {van Leeuwen}, {Walton}, {Arenou}, {Bastian}, {Cropper},
  {Drimmel}, {Katz}, {Lattanzi}, {Bakker}, {Cacciari}, {Casta{\~n}eda},
  {Chaoul}, {Cheek}, {De Angeli}, {Fabricius}, {Guerra}, {Holl}, {Masana},
  {Messineo}, {Mowlavi}, {Nienartowicz}, {Panuzzo}, {Portell}, {Riello},
  {Seabroke}, {Tanga}, {Th{\'e}venin}, {Gracia-Abril}, {Comoretto},
  {Garcia-Reinaldos}, {Teyssier}, {Altmann}, {Andrae}, {Audard},
  {Bellas-Velidis}, {Benson}, {Berthier}, {Blomme}, {Burgess}, {Busso},
  {Carry}, {Cellino}, {Clementini}, {Clotet}, {Creevey}, {Davidson}, {De
  Ridder}, {Delchambre}, {Dell'Oro}, {Ducourant},
  {Fern{\'a}ndez-Hern{\'a}ndez}, {Fouesneau}, {Fr{\'e}mat}, {Galluccio},
  {Garc{\'\i}a-Torres}, {Gonz{\'a}lez-N{\'u}{\~n}ez}, {Gonz{\'a}lez-Vidal},
  {Gosset}, {Guy}, {Halbwachs}, {Hambly}, {Harrison}, {Hern{\'a}ndez},
  {Hestroffer}, {Hodgkin}, {Hutton}, {Jasniewicz}, {Jean-Antoine-Piccolo},
  {Jordan}, {Korn}, {Krone-Martins}, {Lanzafame}, {Lebzelter}, {L{\"o}ffler},
  {Manteiga}, {Marrese}, {Mart{\'\i}n-Fleitas}, {Moitinho}, {Mora}, {Muinonen},
  {Osinde}, {Pancino}, {Pauwels}, {Petit}, {Recio-Blanco}, {Richards},
  {Rimoldini}, {Robin}, {Sarro}, {Siopis}, {Smith}, {Sozzetti}, {S{\"u}veges},
  {Torra}, {van Reeven}, {Abbas}, {Abreu Aramburu}, {Accart}, {Aerts},
  {Altavilla}, {{\'A}lvarez}, {Alvarez}, {Alves}, {Anderson}, {Andrei},
  {Anglada Varela}, {Antiche}, {Antoja}, {Arcay}, {Astraatmadja}, {Bach},
  {Baker}, {Balaguer-N{\'u}{\~n}ez}, {Balm}, {Barache}, {Barata}, {Barbato},
  {Barblan}, {Barklem}, {Barrado}, {Barros}, {Barstow}, {Bartholom{\'e}
  Mu{\~n}oz}, {Bassilana}, {Becciani}, {Bellazzini}, {Berihuete}, {Bertone},
  {Bianchi}, {Bienaym{\'e}}, {Blanco-Cuaresma}, {Boch}, {Boeche}, {Bombrun},
  {Borrachero}, {Bossini}, {Bouquillon}, {Bourda}, {Bragaglia}, {Bramante},
  {Breddels}, {Bressan}, {Brouillet}, {Br{\"u}semeister}, {Brugaletta},
  {Bucciarelli}, {Burlacu}, {Busonero}, {Butkevich}, {Buzzi}, {Caffau},
  {Cancelliere}, {Cannizzaro}, {Cantat-Gaudin}, {Carballo}, {Carlucci},
  {Carrasco}, {Casamiquela}, {Castellani}, {Castro-Ginard}, {Charlot},
  {Chemin}, {Chiavassa}, {Cocozza}, {Costigan}, {Cowell}, {Crifo}, {Crosta},
  {Crowley}, {Cuypers}, {Dafonte}, {Damerdji}, {Dapergolas}, {David}, {David},
  {de Laverny}, {De Luise}, {De March}, {de Martino}, {de Souza}, {de Torres},
  {Debosscher}, {del Pozo}, {Delbo}, {Delgado}, {Delgado}, {Di Matteo},
  {Diakite}, {Diener}, {Distefano}, {Dolding}, {Drazinos}, {Dur{\'a}n},
  {Edvardsson}, {Enke}, {Eriksson}, {Esquej}, {Eynard Bontemps}, {Fabre},
  {Fabrizio}, {Faigler}, {Falc{\~a}o}, {Farr{\`a}s Casas}, {Federici},
  {Fedorets}, {Fernique}, {Figueras}, {Filippi}, {Findeisen}, {Fonti},
  {Fraile}, {Fraser}, {Fr{\'e}zouls}, {Gai}, {Galleti}, {Garabato},
  {Garc{\'\i}a-Sedano}, {Garofalo}, {Garralda}, {Gavel}, {Gavras}, {Gerssen},
  {Geyer}, {Giacobbe}, {Gilmore}, {Girona}, {Giuffrida}, {Glass}, {Gomes},
  {Granvik}, {Gueguen}, {Guerrier}, {Guiraud}, {Guti{\'e}rrez-S{\'a}nchez},
  {Haigron}, {Hatzidimitriou}, {Hauser}, {Haywood}, {Heiter}, {Helmi}, {Heu},
  {Hilger}, {Hobbs}, {Hofmann}, {Holland}, {Huckle}, {Hypki}, {Icardi},
  {Jan{\ss}en}, {Jevardat de Fombelle}, {Jonker}, {Juh{\'a}sz}, {Julbe},
  {Karampelas}, {Kewley}, {Klar}, {Kochoska}, {Kohley}, {Kolenberg},
  {Kontizas}, {Kontizas}, {Koposov}, {Kordopatis}, {Kostrzewa-Rutkowska},
  {Koubsky}, {Lambert}, {Lanza}, {Lasne}, {Lavigne}, {Le Fustec}, {Le
  Poncin-Lafitte}, {Lebreton}, {Leccia}, {Leclerc}, {Lecoeur-Taibi},
  {Lenhardt}, {Leroux}, {Liao}, {Licata}, {Lindstr{\o}m}, {Lister}, {Livanou},
  {Lobel}, {L{\'o}pez}, {Managau}, {Mann}, {Mantelet}, {Marchal}, {Marchant},
  {Marconi}, {Marinoni}, {Marschalk{\'o}}, {Marshall}, {Martino}, {Marton},
  {Mary}, {Massari}, {Matijevi{\v{c}}}, {Mazeh}, {McMillan}, {Messina},
  {Michalik}, {Millar}, {Molina}, {Molinaro}, {Moln{\'a}r}, {Montegriffo},
  {Mor}, {Morbidelli}, {Morel}, {Morris}, {Mulone}, {Muraveva}, {Musella},
  {Nelemans}, {Nicastro}, {Noval}, {O'Mullane}, {Ord{\'e}novic},
  {Ord{\'o}{\~n}ez-Blanco}, {Osborne}, {Pagani}, {Pagano}, {Pailler},
  {Palacin}, {Palaversa}, {Panahi}, {Pawlak}, {Piersimoni}, {Pineau}, {Plachy},
  {Plum}, {Poggio}, {Poujoulet}, {Pr{\v{s}}a}, {Pulone}, {Racero}, {Ragaini},
  {Rambaux}, {Ramos-Lerate}, {Regibo}, {Reyl{\'e}}, {Riclet}, {Ripepi}, {Riva},
  {Rivard}, {Rixon}, {Roegiers}, {Roelens}, {Romero-G{\'o}mez}, {Rowell},
  {Royer}, {Ruiz-Dern}, {Sadowski}, {Sagrist{\`a} Sell{\'e}s}, {Sahlmann},
  {Salgado}, {Salguero}, {Sanna}, {Santana-Ros}, {Sarasso}, {Savietto},
  {Schultheis}, {Sciacca}, {Segol}, {Segovia}, {S{\'e}gransan}, {Shih},
  {Siltala}, {Silva}, {Smart}, {Smith}, {Solano}, {Solitro}, {Sordo}, {Soria
  Nieto}, {Souchay}, {Spagna}, {Spoto}, {Stampa}, {Steele},
  {Steidelm{\"u}ller}, {Stephenson}, {Stoev}, {Suess}, {Surdej}, {Szabados},
  {Szegedi-Elek}, {Tapiador}, {Taris}, {Tauran}, {Taylor}, {Teixeira},
  {Terrett}, {Teyssandier}, {Thuillot}, {Titarenko}, {Torra Clotet}, {Turon},
  {Ulla}, {Utrilla}, {Uzzi}, {Vaillant}, {Valentini}, {Valette}, {van Elteren},
  {Van Hemelryck}, {van Leeuwen}, {Vaschetto}, {Vecchiato}, {Veljanoski},
  {Viala}, {Vicente}, {Vogt}, {von Essen}, {Voss}, {Votruba}, {Voutsinas},
  {Walmsley}, {Weiler}, {Wertz}, {Wevers}, {Wyrzykowski}, {Yoldas},
  {{\v{Z}}erjal}, {Ziaeepour}, {Zorec}, {Zschocke}, {Zucker}, {Zurbach}, \&
  {Zwitter}}]{Brown2018}
{Gaia Collaboration}, {Brown}, A.~G.~A., {Vallenari}, A., {et~al.} 2018, \aap,
  616, A1

\bibitem[{{Gaia Collaboration} {et~al.}(2016){Gaia Collaboration}, {Prusti},
  {de Bruijne}, {Brown}, {Vallenari}, {Babusiaux}, {Bailer-Jones}, {Bastian},
  {Biermann}, {Evans}, {Eyer}, {Jansen}, {Jordi}, {Klioner}, {Lammers},
  {Lindegren}, {Luri}, {Mignard}, {Milligan}, {Panem}, {Poinsignon},
  {Pourbaix}, {Randich}, {Sarri}, {Sartoretti}, {Siddiqui}, {Soubiran},
  {Valette}, {van Leeuwen}, {Walton}, {Aerts}, {Arenou}, {Cropper}, {Drimmel},
  {H{\o}g}, {Katz}, {Lattanzi}, {O'Mullane}, {Grebel}, {Holland}, {Huc},
  {Passot}, {Bramante}, {Cacciari}, {Casta{\~n}eda}, {Chaoul}, {Cheek}, {De
  Angeli}, {Fabricius}, {Guerra}, {Hern{\'a}ndez}, {Jean-Antoine-Piccolo},
  {Masana}, {Messineo}, {Mowlavi}, {Nienartowicz}, {Ord{\'o}{\~n}ez-Blanco},
  {Panuzzo}, {Portell}, {Richards}, {Riello}, {Seabroke}, {Tanga},
  {Th{\'e}venin}, {Torra}, {Els}, {Gracia-Abril}, {Comoretto},
  {Garcia-Reinaldos}, {Lock}, {Mercier}, {Altmann}, {Andrae}, {Astraatmadja},
  {Bellas-Velidis}, {Benson}, {Berthier}, {Blomme}, {Busso}, {Carry},
  {Cellino}, {Clementini}, {Cowell}, {Creevey}, {Cuypers}, {Davidson}, {De
  Ridder}, {de Torres}, {Delchambre}, {Dell'Oro}, {Ducourant}, {Fr{\'e}mat},
  {Garc{\'\i}a-Torres}, {Gosset}, {Halbwachs}, {Hambly}, {Harrison}, {Hauser},
  {Hestroffer}, {Hodgkin}, {Huckle}, {Hutton}, {Jasniewicz}, {Jordan},
  {Kontizas}, {Korn}, {Lanzafame}, {Manteiga}, {Moitinho}, {Muinonen},
  {Osinde}, {Pancino}, {Pauwels}, {Petit}, {Recio-Blanco}, {Robin}, {Sarro},
  {Siopis}, {Smith}, {Smith}, {Sozzetti}, {Thuillot}, {van Reeven}, {Viala},
  {Abbas}, {Abreu Aramburu}, {Accart}, {Aguado}, {Allan}, {Allasia},
  {Altavilla}, {{\'A}lvarez}, {Alves}, {Anderson}, {Andrei}, {Anglada Varela},
  {Antiche}, {Antoja}, {Ant{\'o}n}, {Arcay}, {Atzei}, {Ayache}, {Bach},
  {Baker}, {Balaguer-N{\'u}{\~n}ez}, {Barache}, {Barata}, {Barbier}, {Barblan},
  {Baroni}, {Barrado y Navascu{\'e}s}, {Barros}, {Barstow}, {Becciani},
  {Bellazzini}, {Bellei}, {Bello Garc{\'\i}a}, {Belokurov}, {Bendjoya},
  {Berihuete}, {Bianchi}, {Bienaym{\'e}}, {Billebaud}, {Blagorodnova},
  {Blanco-Cuaresma}, {Boch}, {Bombrun}, {Borrachero}, {Bouquillon}, {Bourda},
  {Bouy}, {Bragaglia}, {Breddels}, {Brouillet}, {Br{\"u}semeister},
  {Bucciarelli}, {Budnik}, {Burgess}, {Burgon}, {Burlacu}, {Busonero}, {Buzzi},
  {Caffau}, {Cambras}, {Campbell}, {Cancelliere}, {Cantat-Gaudin}, {Carlucci},
  {Carrasco}, {Castellani}, {Charlot}, {Charnas}, {Charvet}, {Chassat},
  {Chiavassa}, {Clotet}, {Cocozza}, {Collins}, {Collins}, {Costigan}, {Crifo},
  {Cross}, {Crosta}, {Crowley}, {Dafonte}, {Damerdji}, {Dapergolas}, {David},
  {David}, {De Cat}, {de Felice}, {de Laverny}, {De Luise}, {De March}, {de
  Martino}, {de Souza}, {Debosscher}, {del Pozo}, {Delbo}, {Delgado},
  {Delgado}, {di Marco}, {Di Matteo}, {Diakite}, {Distefano}, {Dolding}, {Dos
  Anjos}, {Drazinos}, {Dur{\'a}n}, {Dzigan}, {Ecale}, {Edvardsson}, {Enke},
  {Erdmann}, {Escolar}, {Espina}, {Evans}, {Eynard Bontemps}, {Fabre},
  {Fabrizio}, {Faigler}, {Falc{\~a}o}, {Farr{\`a}s Casas}, {Faye}, {Federici},
  {Fedorets}, {Fern{\'a}ndez-Hern{\'a}ndez}, {Fernique}, {Fienga}, {Figueras},
  {Filippi}, {Findeisen}, {Fonti}, {Fouesneau}, {Fraile}, {Fraser}, {Fuchs},
  {Furnell}, {Gai}, {Galleti}, {Galluccio}, {Garabato}, {Garc{\'\i}a-Sedano},
  {Gar{\'e}}, {Garofalo}, {Garralda}, {Gavras}, {Gerssen}, {Geyer}, {Gilmore},
  {Girona}, {Giuffrida}, {Gomes}, {Gonz{\'a}lez-Marcos},
  {Gonz{\'a}lez-N{\'u}{\~n}ez}, {Gonz{\'a}lez-Vidal}, {Granvik}, {Guerrier},
  {Guillout}, {Guiraud}, {G{\'u}rpide}, {Guti{\'e}rrez-S{\'a}nchez}, {Guy},
  {Haigron}, {Hatzidimitriou}, {Haywood}, {Heiter}, {Helmi}, {Hobbs},
  {Hofmann}, {Holl}, {Holland}, {Hunt}, {Hypki}, {Icardi}, {Irwin}, {Jevardat
  de Fombelle}, {Jofr{\'e}}, {Jonker}, {Jorissen}, {Julbe}, {Karampelas},
  {Kochoska}, {Kohley}, {Kolenberg}, {Kontizas}, {Koposov}, {Kordopatis},
  {Koubsky}, {Kowalczyk}, {Krone-Martins}, {Kudryashova}, {Kull}, {Bachchan},
  {Lacoste-Seris}, {Lanza}, {Lavigne}, {Le Poncin-Lafitte}, {Lebreton},
  {Lebzelter}, {Leccia}, {Leclerc}, {Lecoeur-Taibi}, {Lemaitre}, {Lenhardt},
  {Leroux}, {Liao}, {Licata}, {Lindstr{\o}m}, {Lister}, {Livanou}, {Lobel},
  {L{\"o}ffler}, {L{\'o}pez}, {Lopez-Lozano}, {Lorenz}, {Loureiro},
  {MacDonald}, {Magalh{\~a}es Fernandes}, {Managau}, {Mann}, {Mantelet},
  {Marchal}, {Marchant}, {Marconi}, {Marie}, {Marinoni}, {Marrese},
  {Marschalk{\'o}}, {Marshall}, {Mart{\'\i}n-Fleitas}, {Martino}, {Mary},
  {Matijevi{\v{c}}}, {Mazeh}, {McMillan}, {Messina}, {Mestre}, {Michalik},
  {Millar}, {Miranda}, {Molina}, {Molinaro}, {Molinaro}, {Moln{\'a}r},
  {Moniez}, {Montegriffo}, {Monteiro}, {Mor}, {Mora}, {Morbidelli}, {Morel},
  {Morgenthaler}, {Morley}, {Morris}, {Mulone}, {Muraveva}, {Musella},
  {Narbonne}, {Nelemans}, {Nicastro}, {Noval}, {Ord{\'e}novic},
  {Ordieres-Mer{\'e}}, {Osborne}, {Pagani}, {Pagano}, {Pailler}, {Palacin},
  {Palaversa}, {Parsons}, {Paulsen}, {Pecoraro}, {Pedrosa}, {Pentik{\"a}inen},
  {Pereira}, {Pichon}, {Piersimoni}, {Pineau}, {Plachy}, {Plum}, {Poujoulet},
  {Pr{\v{s}}a}, {Pulone}, {Ragaini}, {Rago}, {Rambaux}, {Ramos-Lerate},
  {Ranalli}, {Rauw}, {Read}, {Regibo}, {Renk}, {Reyl{\'e}}, {Ribeiro},
  {Rimoldini}, {Ripepi}, {Riva}, {Rixon}, {Roelens}, {Romero-G{\'o}mez},
  {Rowell}, {Royer}, {Rudolph}, {Ruiz-Dern}, {Sadowski}, {Sagrist{\`a}
  Sell{\'e}s}, {Sahlmann}, {Salgado}, {Salguero}, {Sarasso}, {Savietto},
  {Schnorhk}, {Schultheis}, {Sciacca}, {Segol}, {Segovia}, {Segransan},
  {Serpell}, {Shih}, {Smareglia}, {Smart}, {Smith}, {Solano}, {Solitro},
  {Sordo}, {Soria Nieto}, {Souchay}, {Spagna}, {Spoto}, {Stampa}, {Steele},
  {Steidelm{\"u}ller}, {Stephenson}, {Stoev}, {Suess}, {S{\"u}veges}, {Surdej},
  {Szabados}, {Szegedi-Elek}, {Tapiador}, {Taris}, {Tauran}, {Taylor},
  {Teixeira}, {Terrett}, {Tingley}, {Trager}, {Turon}, {Ulla}, {Utrilla},
  {Valentini}, {van Elteren}, {Van Hemelryck}, {van Leeuwen}, {Varadi},
  {Vecchiato}, {Veljanoski}, {Via}, {Vicente}, {Vogt}, {Voss}, {Votruba},
  {Voutsinas}, {Walmsley}, {Weiler}, {Weingrill}, {Werner}, {Wevers},
  {Whitehead}, {Wyrzykowski}, {Yoldas}, {{\v{Z}}erjal}, {Zucker}, {Zurbach},
  {Zwitter}, {Alecu}, {Allen}, {Allende Prieto}, {Amorim},
  {Anglada-Escud{\'e}}, {Arsenijevic}, {Azaz}, {Balm}, {Beck}, {Bernstein},
  {Bigot}, {Bijaoui}, {Blasco}, {Bonfigli}, {Bono}, {Boudreault}, {Bressan},
  {Brown}, {Brunet}, {Bunclark}, {Buonanno}, {Butkevich}, {Carret}, {Carrion},
  {Chemin}, {Ch{\'e}reau}, {Corcione}, {Darmigny}, {de Boer}, {de Teodoro}, {de
  Zeeuw}, {Delle Luche}, {Domingues}, {Dubath}, {Fodor}, {Fr{\'e}zouls},
  {Fries}, {Fustes}, {Fyfe}, {Gallardo}, {Gallegos}, {Gardiol}, {Gebran},
  {Gomboc}, {G{\'o}mez}, {Grux}, {Gueguen}, {Heyrovsky}, {Hoar}, {Iannicola},
  {Isasi Parache}, {Janotto}, {Joliet}, {Jonckheere}, {Keil}, {Kim},
  {Klagyivik}, {Klar}, {Knude}, {Kochukhov}, {Kolka}, {Kos}, {Kutka}, {Lainey},
  {LeBouquin}, {Liu}, {Loreggia}, {Makarov}, {Marseille}, {Martayan},
  {Martinez-Rubi}, {Massart}, {Meynadier}, {Mignot}, {Munari}, {Nguyen},
  {Nordlander}, {Ocvirk}, {O'Flaherty}, {Olias Sanz}, {Ortiz}, {Osorio},
  {Oszkiewicz}, {Ouzounis}, {Palmer}, {Park}, {Pasquato}, {Peltzer}, {Peralta},
  {P{\'e}turaud}, {Pieniluoma}, {Pigozzi}, {Poels}, {Prat}, {Prod'homme},
  {Raison}, {Rebordao}, {Risquez}, {Rocca-Volmerange}, {Rosen}, {Ruiz-Fuertes},
  {Russo}, {Sembay}, {Serraller Vizcaino}, {Short}, {Siebert}, {Silva},
  {Sinachopoulos}, {Slezak}, {Soffel}, {Sosnowska}, {Strai{\v{z}}ys}, {ter
  Linden}, {Terrell}, {Theil}, {Tiede}, {Troisi}, {Tsalmantza}, {Tur},
  {Vaccari}, {Vachier}, {Valles}, {Van Hamme}, {Veltz}, {Virtanen}, {Wallut},
  {Wichmann}, {Wilkinson}, {Ziaeepour}, \& {Zschocke}}]{Prusti2016}
{Gaia Collaboration}, {Prusti}, T., {de Bruijne}, J.~H.~J., {et~al.} 2016,
  \aap, 595, A1

\bibitem[{{Ginzburg} \& {Chiang}(2019)}]{Ginzburg2019}
{Ginzburg}, S. \& {Chiang}, E. 2019, \mnras, 490, 4334

\bibitem[{{Haffert} {et~al.}(2019){Haffert}, {Bohn}, {de Boer}, {Snellen},
  {Brinchmann}, {Girard}, {Keller}, \& {Bacon}}]{Haffert2019}
{Haffert}, S.~Y., {Bohn}, A.~J., {de Boer}, J., {et~al.} 2019, Nature
  Astronomy, 3, 749

\bibitem[{{Hashimoto} {et~al.}(2012){Hashimoto}, {Dong}, {Kudo}, {Honda},
  {McClure}, {Zhu}, {Muto}, {Wisniewski}, {Abe}, {Brandner}, {Brandt},
  {Carson}, {Egner}, {Feldt}, {Fukagawa}, {Goto}, {Grady}, {Guyon}, {Hayano},
  {Hayashi}, {Hayashi}, {Henning}, {Hodapp}, {Ishii}, {Iye}, {Janson},
  {Kandori}, {Knapp}, {Kusakabe}, {Kuzuhara}, {Kwon}, {Matsuo}, {Mayama},
  {McElwain}, {Miyama}, {Morino}, {Moro-Martin}, {Nishimura}, {Pyo}, {Serabyn},
  {Suenaga}, {Suto}, {Suzuki}, {Takahashi}, {Takami}, {Takato}, {Terada},
  {Thalmann}, {Tomono}, {Turner}, {Watanabe}, {Yamada}, {Takami}, {Usuda}, \&
  {Tamura}}]{Hashimoto2012}
{Hashimoto}, J., {Dong}, R., {Kudo}, T., {et~al.} 2012, \apjl, 758, L19

\bibitem[{{Isella} {et~al.}(2019){Isella}, {Benisty}, {Teague}, {Bae},
  {Keppler}, {Facchini}, \& {P{\'e}rez}}]{Isella2019}
{Isella}, A., {Benisty}, M., {Teague}, R., {et~al.} 2019, \apjl, 879, L25

\bibitem[{{Isella} {et~al.}(2014){Isella}, {Chandler}, {Carpenter},
  {P{\'e}rez}, \& {Ricci}}]{Isella2014}
{Isella}, A., {Chandler}, C.~J., {Carpenter}, J.~M., {P{\'e}rez}, L.~M., \&
  {Ricci}, L. 2014, \apj, 788, 129

\bibitem[{{Keppler} {et~al.}(2018){Keppler}, {Benisty}, {M{\"u}ller},
  {Henning}, {van Boekel}, {Cantalloube}, {Ginski}, {van Holstein}, {Maire},
  {Pohl}, {Samland }, {Avenhaus}, {Baudino}, {Boccaletti}, {de Boer},
  {Bonnefoy}, {Chauvin}, {Desidera}, {Langlois}, {Lazzoni}, {Marleau},
  {Mordasini}, {Pawellek}, {Stolker}, {Vigan}, {Zurlo}, {Birnstiel},
  {Brandner}, {Feldt}, {Flock}, {Girard}, {Gratton}, {Hagelberg}, {Isella},
  {Janson}, {Juhasz}, {Kemmer}, {Kral}, {Lagrange}, {Launhardt}, {Matter},
  {M{\'e}nard}, {Milli}, {Molli{\`e}re}, {Olofsson}, {P{\'e}rez}, {Pinilla},
  {Pinte}, {Quanz}, {Schmidt}, {Udry}, {Wahhaj}, {Williams}, {Buenzli},
  {Cudel}, {Dominik}, {Galicher}, {Kasper}, {Lannier}, {Mesa}, {Mouillet},
  {Peretti}, {Perrot}, {Salter}, {Sissa}, {Wildi}, {Abe}, {Antichi},
  {Augereau}, {Baruffolo}, {Baudoz}, {Bazzon}, {Beuzit}, {Blanchard}, {Brems},
  {Buey}, {De Caprio}, {Carbillet}, {Carle}, {Cascone}, {Cheetham}, {Claudi},
  {Costille}, {Delboulb{\'e}}, {Dohlen}, {Fantinel}, {Feautrier}, {Fusco},
  {Giro}, {Gluck}, {Gry}, {Hubin}, {Hugot}, {Jaquet}, {Le Mignant}, {Llored},
  {Madec}, {Magnard}, {Martinez}, {Maurel}, {Meyer}, {M{\"o}ller-Nilsson},
  {Moulin}, {Mugnier}, {Orign{\'e}}, {Pavlov}, {Perret}, {Petit}, {Pragt},
  {Puget}, {Rabou}, {Ramos}, {Rigal}, {Rochat}, {Roelfsema}, {Rousset}, {Roux},
  {Salasnich}, {Sauvage}, {Sevin}, {Soenke}, {Stadler}, {Suarez}, {Turatto}, \&
  {Weber}}]{Keppler2018}
{Keppler}, M., {Benisty}, M., {M{\"u}ller}, A., {et~al.} 2018, \aap, 617, A44

\bibitem[{{Keppler} {et~al.}(2019){Keppler}, {Teague}, {Bae}, {Benisty},
  {Henning}, {van Boekel}, {Chapillon}, {Pinilla}, {Williams}, {Bertrang},
  {Facchini}, {Flock}, {Ginski}, {Juhasz}, {Klahr}, {Liu}, {M{\"u}ller},
  {P{\'e}rez}, {Pohl}, {Rosotti}, {Samland}, \& {Semenov}}]{Keppler2019}
{Keppler}, M., {Teague}, R., {Bae}, J., {et~al.} 2019, \aap, 625, A118

\bibitem[{{Kley}(1999)}]{Kley1999}
{Kley}, W. 1999, \mnras, 303, 696

\bibitem[{{Kraus} \& {Ireland}(2012)}]{Kraus2012}
{Kraus}, A.~L. \& {Ireland}, M.~J. 2012, \apj, 745, 5

\bibitem[{{Long} {et~al.}(2018){Long}, {Akiyama}, {Sitko}, {Fernandes},
  {Assani}, {Grady}, {Cure}, {Danchi}, {Dong}, {Fukagawa}, {Hasegawa},
  {Hashimoto}, {Henning}, {Inutsuka}, {Kraus}, {Kwon}, {Lisse}, {Baobabu Liu},
  {Mayama}, {Muto}, {Nakagawa}, {Takami}, {Tamura}, {Currie}, {Wisniewski}, \&
  {Yang}}]{Long2018}
{Long}, Z.~C., {Akiyama}, E., {Sitko}, M., {et~al.} 2018, \apj, 858, 112

\bibitem[{{Lubow} {et~al.}(1999){Lubow}, {Seibert}, \&
  {Artymowicz}}]{Lubow1999}
{Lubow}, S.~H., {Seibert}, M., \& {Artymowicz}, P. 1999, \apj, 526, 1001

\bibitem[{{Mesa} {et~al.}(2019){Mesa}, {Keppler}, {Cantalloube}, {Rodet},
  {Charnay}, {Gratton}, {Langlois}, {Boccaletti}, {Bonnefoy}, {Vigan},
  {Flasseur}, {Bae}, {Benisty}, {Chauvin}, {de Boer}, {Desidera}, {Henning},
  {Lagrange}, {Meyer}, {Milli}, {M{\"u}ller}, {Pairet}, {Zurlo}, {Antoniucci},
  {Baudino}, {Brown Sevilla}, {Cascone}, {Cheetham}, {Claudi}, {Delorme},
  {D'Orazi}, {Feldt}, {Hagelberg}, {Janson}, {Kral}, {Lagadec}, {Lazzoni},
  {Ligi}, {Maire}, {Martinez}, {Menard}, {Meunier}, {Perrot}, {Petrus},
  {Pinte}, {Rickman}, {Rochat}, {Rouan}, {Samland}, {Sauvage}, {Schmidt},
  {Udry}, {Weber}, \& {Wildi}}]{Mesa2019}
{Mesa}, D., {Keppler}, M., {Cantalloube}, F., {et~al.} 2019, \aap, 632, A25

\bibitem[{{Min} {et~al.}(2012){Min}, {Canovas}, {Mulders}, \&
  {Keller}}]{Min2012}
{Min}, M., {Canovas}, H., {Mulders}, G.~D., \& {Keller}, C.~U. 2012, \aap, 537,
  A75

\bibitem[{{Min} {et~al.}(2009){Min}, {Dullemond}, {Dominik}, {de Koter}, \&
  {Hovenier}}]{Min2009}
{Min}, M., {Dullemond}, C.~P., {Dominik}, C., {de Koter}, A., \& {Hovenier},
  J.~W. 2009, \aap, 497, 155

\bibitem[{{M{\"u}ller} {et~al.}(2018){M{\"u}ller}, {Keppler}, {Henning},
  {Samland}, {Chauvin}, {Beust}, {Maire}, {Molaverdikhani}, {van Boekel},
  {Benisty}, {Boccaletti}, {Bonnefoy}, {Cantalloube}, {Charnay}, {Baudino},
  {Gennaro}, {Long}, {Cheetham}, {Desidera}, {Feldt}, {Fusco}, {Girard},
  {Gratton}, {Hagelberg}, {Janson}, {Lagrange}, {Langlois}, {Lazzoni}, {Ligi},
  {M{\'e}nard}, {Mesa}, {Meyer}, {Molli{\`e}re}, {Mordasini}, {Moulin},
  {Pavlov}, {Pawellek}, {Quanz}, {Ramos}, {Rouan}, {Sissa}, {Stadler}, {Vigan},
  {Wahhaj}, {Weber}, \& {Zurlo}}]{Muller2018}
{M{\"u}ller}, A., {Keppler}, M., {Henning}, T., {et~al.} 2018, \aap, 617, L2

\bibitem[{{Muro-Arena} {et~al.}(2018){Muro-Arena}, {Dominik}, {Waters}, {Min},
  {Klarmann}, {Ginski}, {Isella}, {Benisty}, {Pohl}, {Garufi}, {Hagelberg},
  {Langlois}, {Menard}, {Pinte}, {Sezestre}, {van der Plas}, {Villenave},
  {Delboulb{\'e}}, {Magnard}, {M{\"o}ller-Nilsson}, {Pragt}, {Rabou}, \&
  {Roelfsema}}]{Muro2018}
{Muro-Arena}, G.~A., {Dominik}, C., {Waters}, L.~B.~F.~M., {et~al.} 2018, \aap,
  614, A24

\bibitem[{{Pinilla} {et~al.}(2015){Pinilla}, {de Juan Ovelar}, {Ataiee},
  {Benisty}, {Birnstiel}, {van Dishoeck}, \& {Min}}]{Pinilla2015}
{Pinilla}, P., {de Juan Ovelar}, M., {Ataiee}, S., {et~al.} 2015, \aap, 573, A9

\bibitem[{{Pinilla} {et~al.}(2016){Pinilla}, {Flock}, {Ovelar}, \&
  {Birnstiel}}]{Pinilla2016}
{Pinilla}, P., {Flock}, M., {Ovelar}, M. d.~J., \& {Birnstiel}, T. 2016, \aap,
  596, A81

\bibitem[{{Pinte} {et~al.}(2016){Pinte}, {Dent}, {M{\'e}nard}, {Hales}, {Hill},
  {Cortes}, \& {de Gregorio-Monsalvo}}]{Pinte2016}
{Pinte}, C., {Dent}, W.~R.~F., {M{\'e}nard}, F., {et~al.} 2016, \apj, 816, 25

\bibitem[{{Quillen} \& {Trilling}(1998)}]{Quillen1998}
{Quillen}, A.~C. \& {Trilling}, D.~E. 1998, \apj, 508, 707

\bibitem[{{Rab} {et~al.}(2020){Rab}, {Kamp}, {Dominik}, {Ginski}, {Muro-Arena},
  {Thi}, {Waters}, \& {Woitke}}]{Rab2020}
{Rab}, C., {Kamp}, I., {Dominik}, C., {et~al.} 2020, \aap, 642, A165

\bibitem[{{Rab} {et~al.}(2019){Rab}, {Kamp}, {Ginski}, {Oberg}, {Muro-Arena},
  {Dominik}, {Waters}, {Thi}, \& {Woitke}}]{Rab2019}
{Rab}, C., {Kamp}, I., {Ginski}, C., {et~al.} 2019, \aap, 624, A16

\bibitem[{{Riaud} {et~al.}(2006){Riaud}, {Mawet}, {Absil}, {Boccaletti},
  {Baudoz}, {Herwats}, \& {Surdej}}]{Riaud2006}
{Riaud}, P., {Mawet}, D., {Absil}, O., {et~al.} 2006, \aap, 458, 317

\bibitem[{{Sasaki} {et~al.}(2010){Sasaki}, {Stewart}, \& {Ida}}]{Sasaki2010}
{Sasaki}, T., {Stewart}, G.~R., \& {Ida}, S. 2010, \apj, 714, 1052

\bibitem[{{Schmid} {et~al.}(2006){Schmid}, {Joos}, \& {Tschan}}]{Schmid2006}
{Schmid}, H.~M., {Joos}, F., \& {Tschan}, D. 2006, \aap, 452, 657

\bibitem[{{Shakura} \& {Sunyaev}(1973)}]{Shakura1973}
{Shakura}, N.~I. \& {Sunyaev}, R.~A. 1973, \aap, 500, 33

\bibitem[{{Szul{\'a}gyi} {et~al.}(2019){Szul{\'a}gyi}, {Dullemond}, {Pohl}, \&
  {Quanz}}]{Szulagyi2019}
{Szul{\'a}gyi}, J., {Dullemond}, C.~P., {Pohl}, A., \& {Quanz}, S.~P. 2019,
  \mnras, 487, 1248

\bibitem[{{Szul{\'a}gyi} {et~al.}(2017){Szul{\'a}gyi}, {Mayer}, \&
  {Quinn}}]{Szulagyi2017}
{Szul{\'a}gyi}, J., {Mayer}, L., \& {Quinn}, T. 2017, \mnras, 464, 3158

\bibitem[{Teague(2019)}]{Teague2019}
Teague, R. 2019, The Journal of Open Source Software, 4, 1632

\bibitem[{{Toci} {et~al.}(2020){Toci}, {Lodato}, {Christiaens}, {Fedele},
  {Pinte}, {Price}, \& {Testi}}]{Toci2020}
{Toci}, C., {Lodato}, G., {Christiaens}, V., {et~al.} 2020, \mnras, 499, 2015

\bibitem[{{Vasyunin} {et~al.}(2011){Vasyunin}, {Wiebe}, {Birnstiel},
  {Zhukovska}, {Henning}, \& {Dullemond}}]{Vasyunin2011}
{Vasyunin}, A.~I., {Wiebe}, D.~S., {Birnstiel}, T., {et~al.} 2011, \apj, 727,
  76

\bibitem[{{Villenave} {et~al.}(2019){Villenave}, {Benisty}, {Dent},
  {M{\'e}nard}, {Garufi}, {Ginski}, {Pinilla}, {Pinte}, {Williams}, {de Boer},
  {Morino}, {Fukagawa}, {Dominik}, {Flock}, {Henning}, {Juh{\'a}sz}, {Keppler},
  {Muro-Arena}, {Olofsson}, {P{\'e}rez}, {van der Plas}, {Zurlo}, {Carle},
  {Feautrier}, {Pavlov}, {Pragt}, {Ramos}, {Sauvage}, {Stadler}, \&
  {Weber}}]{Villenave2019}
{Villenave}, M., {Benisty}, M., {Dent}, W.~R.~F., {et~al.} 2019, \aap, 624, A7

\bibitem[{{Wang} {et~al.}(2020){Wang}, {Ginzburg}, {Ren}, {Wallack}, {Gao},
  {Mawet}, {Bond}, {Cetre}, {Wizinowich}, {De Rosa}, {Ruane}, {Liu}, {Absil},
  {Alvarez}, {Baranec}, {Choquet}, {Chun}, {Defr{\`e}re}, {Delorme},
  {Duch{\^e}ne}, {Forsberg}, {Ghez}, {Guyon}, {Hall}, {Huby}, {Jolivet},
  {Jensen-Clem}, {Jovanovic}, {Karlsson}, {Lilley}, {Matthews}, {M{\'e}nard},
  {Meshkat}, {Millar-Blanchaer}, {Ngo}, {Orban de Xivry}, {Pinte}, {Ragland},
  {Serabyn}, {Catal{\'a}n}, {Wang}, {Wetherell}, {Williams}, {Ygouf}, \&
  {Zuckerman}}]{Wang2020}
{Wang}, J.~J., {Ginzburg}, S., {Ren}, B., {et~al.} 2020, \aj, 159, 263

\bibitem[{{Wang} {et~al.}(2021){Wang}, {Vigan}, {Lacour}, {Nowak}, {Stolker},
  {De Rosa}, {Ginzburg}, {Gao}, {Abuter}, {Amorim}, {Asensio-Torres},
  {Baub{\"o}ck}, {Benisty}, {Berger}, {Beust}, {Beuzit}, {Blunt}, {Boccaletti},
  {Bohn}, {Bonnefoy}, {Bonnet}, {Brandner}, {Cantalloube}, {Caselli},
  {Charnay}, {Chauvin}, {Choquet}, {Christiaens}, {Cl{\'e}net}, {Coud{\'e} Du
  Foresto}, {Cridland}, {de Zeeuw}, {Dembet}, {Dexter}, {Drescher}, {Duvert},
  {Eckart}, {Eisenhauer}, {Facchini}, {Gao}, {Garcia}, {Garcia Lopez},
  {Gardner}, {Gendron}, {Genzel}, {Gillessen}, {Girard}, {Haubois},
  {Hei{\ss}el}, {Henning}, {Hinkley}, {Hippler}, {Horrobin}, {Houll{\'e}},
  {Hubert}, {Jim{\'e}nez-Rosales}, {Jocou}, {Kammerer}, {Keppler}, {Kervella},
  {Meyer}, {Kreidberg}, {Lagrange}, {Lapeyr{\`e}re}, {Le Bouquin}, {L{\'e}na},
  {Lutz}, {Maire}, {M{\'e}nard}, {M{\'e}rand}, {Molli{\`e}re}, {Monnier},
  {Mouillet}, {M{\"u}ller}, {Nasedkin}, {Ott}, {Otten}, {Paladini}, {Paumard},
  {Perraut}, {Perrin}, {Pfuhl}, {Pueyo}, {Rameau}, {Rodet},
  {Rodr{\'\i}guez-Coira}, {Rousset}, {Scheithauer}, {Shangguan}, {Shimizu},
  {Stadler}, {Straub}, {Straubmeier}, {Sturm}, {Tacconi}, {van Dishoeck},
  {Vincent}, {von Fellenberg}, {Ward-Duong}, {Widmann}, {Wieprecht},
  {Wiezorrek}, {Woillez}, \& {Gravity Collaboration}}]{Wang2021}
{Wang}, J.~J., {Vigan}, A., {Lacour}, S., {et~al.} 2021, \aj, 161, 148

\bibitem[{{Welch} {et~al.}(2004){Welch}, {Webster}, {Mundy}, {Volgenau}, \&
  {Looney}}]{Welch2004}
{Welch}, W.~J., {Webster}, Z., {Mundy}, L., {Volgenau}, N., \& {Looney}, L.
  2004, in IAU Symposium, Vol. 213, Bioastronomy 2002: Life Among the Stars,
  ed. R.~{Norris} \& F.~{Stootman}, 59

\bibitem[{{Woitke}(2015)}]{W2015}
{Woitke}, P. 2015, in European Physical Journal Web of Conferences, Vol. 102,
  European Physical Journal Web of Conferences, 00007

\bibitem[{{Woitke} {et~al.}(2016){Woitke}, {Min}, {Pinte}, {Thi}, {Kamp},
  {Rab}, {Anthonioz}, {Antonellini}, {Baldovin-Saavedra}, {Carmona}, {Dominik},
  {Dionatos}, {Greaves}, {G{\"u}del}, {Ilee}, {Liebhart}, {M{\'e}nard},
  {Rigon}, {Waters}, {Aresu}, {Meijerink}, \& {Spaans}}]{Woitke2016}
{Woitke}, P., {Min}, M., {Pinte}, C., {et~al.} 2016, \aap, 586, A103

\bibitem[{{Zhu}(2015)}]{Zhu2015}
{Zhu}, Z. 2015, \apj, 799, 16

\bibitem[{{Zhu} {et~al.}(2019){Zhu}, {Zhang}, {Jiang}, {Kataoka}, {Birnstiel},
  {Dullemond}, {Andrews}, {Huang}, {P{\'e}rez}, {Carpenter}, {Bai}, {Wilner},
  \& {Ricci}}]{Zhu2019}
{Zhu}, Z., {Zhang}, S., {Jiang}, Y.-F., {et~al.} 2019, \apjl, 877, L18

\end{thebibliography}

\begin{appendix}

\onecolumn
\section{Physical structure of the modelled protoplanetary disk around PDS 70.}
\label{sect:2D_structure}
\begin{figure*}[h!]
\begin{subfigure}{0.5\textwidth}
  \centering
  \includegraphics[width=1.0\linewidth]{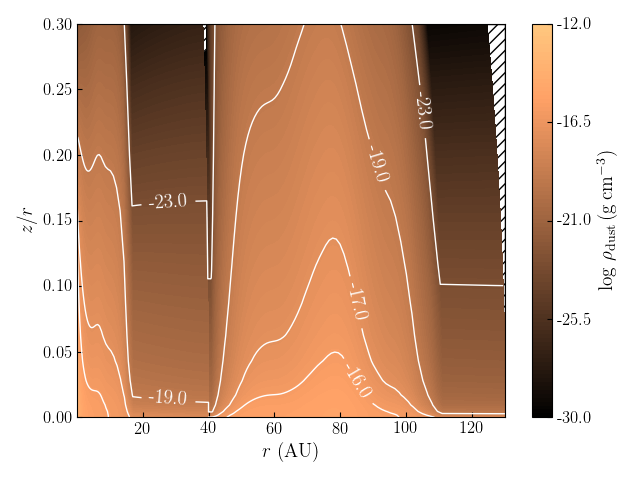}
  \label{fig:sfig1}
\end{subfigure}%
\begin{subfigure}{0.5\textwidth}
  \centering
  \includegraphics[width=1.0\linewidth]{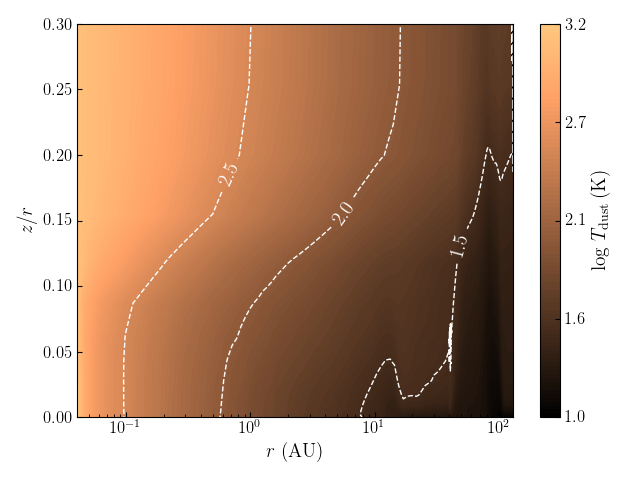}
  \label{fig:sfig2}
\end{subfigure}
\caption{\textit{Left:} Vertical cut in the density structure of the protoplanetary disk. Labels along the contours represent the logarithm of the density. The patched area shows the locations in the disk where $\log \rho_\mathrm{dust}<-30$. \textit{Right:} Vertical cut of the equilibrium temperature field. Labels along the dashed contours show the logarithm of the temperature.}
\label{fig:fig}
\end{figure*}

\begin{figure*}[h!]
\centering
\includegraphics[width=1.0\linewidth]{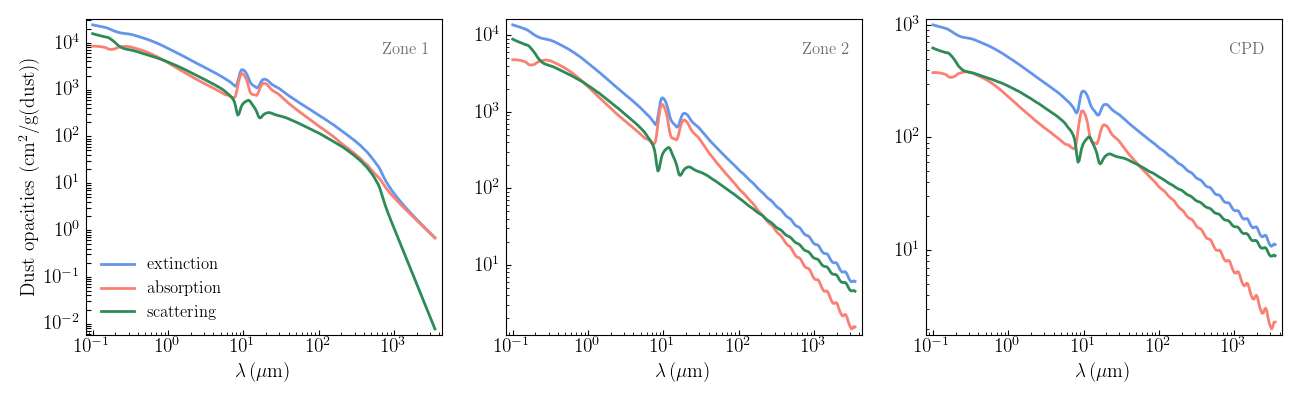}
\caption{\small Extinction, absorption, and scattering opacities computed for the dust grains populating each zone of the model. The wiggles observed at $\lambda>300 \, \microns$ are a numerical artefact product of the implementation of a continuous size distribution in the form of discrete bins.}
\label{fig:opacities}
\end{figure*}

\begin{multicols}{2}
In this section we show the vertical cut in the three-dimensional structure of some relevant quantities of our model for the PDS 70 disk. In Fig. \ref{fig:fig} we report the cuts in the density structure and the equilibrium temperature. In the absence of a planetary companion or any other localised substructure, these fields are azimuthally symmetric. The dust opacities for each zone are shown in Fig. \ref{fig:opacities}.
\end{multicols}

\end{appendix}

\end{document}